\documentclass[12pt]{aastex6}

\usepackage{graphicx,amssymb,amsmath,times}
\graphicspath{{../}{../PLW/}{../figures2/}{../temp/oldcatalogs/}{../temp/170708_catalogs/}}
\usepackage{longtable}
\usepackage{bm} 
\usepackage{url}
\usepackage[varg]{txfonts}
\usepackage[T1]{fontenc}
\usepackage[latin9]{inputenc}
\usepackage{natbib}
\def\simlt{\lower.5ex\hbox{\ltsima}}
\def\simgt{\lower.5ex\hbox{\gtsima}}

\def\kpc{{\rm\,kpc}}

\def\gtsim{\;\lower.6ex\hbox{$\sim$}\kern-6.7pt\raise.4ex\hbox{$>$}\;}
\def\ltsim{\;\lower.6ex\hbox{$\sim$}\kern-6.9pt\raise.4ex\hbox{$<$}\;}

\def\min{${}^{\prime}$}
\def\sec{${}^{\prime\prime}$}
\def\hst{{\it HST\/}}
\def\bmv{\hbox{\it B--V\/}}

\def\bmi{\hbox{\it B--I\/}}

\def\ngc#1{\hbox{NGC$\,$#1}}

\shorttitle{$\omega$ Centauri NIR $JHK_s$ Photometry}
\shortauthors{Braga et~al.}

\begin{document}

\title{On the RR Lyrae stars in globulars: V. the complete Near-Infrared ($JHK_s$) census of 
$\omega$ Centauri RR Lyrae variables\footnote{{This publication makes use of data gathered 
with the Magellan/Baade Telescope at Las Campanas Observatory, the Blanco Telescope
at Cerro Tololo Inter-American Observatory, NTT at La Silla 
(ESO Program IDs: 64.N-0038(A), 66.D-0557(A), 
68.D-0545(A), 073.D-0313(A), ID 073.D-0313(A) and 59.A-9004(D)), 
VISTA at Paranal (ESO Program ID: 179.A-2010) and VLT at 
Paranal (ESO Program ID: ID96406).}}}


\author{V.~F.~Braga\altaffilmark{1,2,3,4}, 
P.~B.~Stetson\altaffilmark{5},
G.~Bono\altaffilmark{3,6},
M.~Dall'Ora\altaffilmark{7}, 
I.~Ferraro\altaffilmark{6}, 
G.~Fiorentino\altaffilmark{8},
G.~Iannicola\altaffilmark{6}, 
M.~Marconi\altaffilmark{7}, 
M.~Marengo\altaffilmark{9}, 
A.~J.~Monson\altaffilmark{10},
J.~Neeley\altaffilmark{9}, 
S.~E.~Persson\altaffilmark{10},
R.~L.~Beaton\altaffilmark{10},
R.~Buonanno\altaffilmark{3,11}, 
A.~Calamida\altaffilmark{12}, 
M.~Castellani\altaffilmark{6}, 
E.~Di Carlo\altaffilmark{11},
M.~Fabrizio\altaffilmark{4},
W.~L.~Freedman\altaffilmark{13},
L.~Inno\altaffilmark{14}, 
B.~F.~Madore\altaffilmark{10},
D.~Magurno\altaffilmark{3}, 
E.~Marchetti\altaffilmark{15}, 
S.~Marinoni\altaffilmark{4},
P.~Marrese\altaffilmark{4},
N.~Matsunaga\altaffilmark{16},
D.~Minniti\altaffilmark{1,2,17}, 
M.~Monelli\altaffilmark{18},
M.~Nonino\altaffilmark{19},
A.~M.~Piersimoni\altaffilmark{11},
A.~Pietrinferni\altaffilmark{11},
P.~Prada-Moroni\altaffilmark{20,21},
L.~Pulone\altaffilmark{6},
R.~Stellingwerf\altaffilmark{22},
E.~Tognelli\altaffilmark{20,21},
A.~R.~Walker\altaffilmark{23}
E.~Valenti\altaffilmark{15},
M.~Zoccali\altaffilmark{1,24} 
}
\altaffiltext{1}{Instituto Milenio de Astrofisica, Santiago, Chile}
\altaffiltext{2}{Departamento de Fisica, Facultad de Ciencias Exactas, Universidad Andres Bello, Fernandez Concha 700, Las Condes, Santiago, Chile}
\altaffiltext{3}{Department of Physics, Universit\`a di Roma Tor Vergata, via della Ricerca Scientifica 1, 00133 Roma, Italy}
\altaffiltext{4}{SSDC, via del Politecnico snc, 00133 Roma, Italy}
\altaffiltext{5}{NRC-Herzberg, Dominion Astrophysical Observatory, 5071 West Saanich Road, Victoria BC V9E 2E7, Canada}
\altaffiltext{6}{INAF-Osservatorio Astronomico di Roma, via Frascati 33, 00040 Monte Porzio Catone, Italy}
\altaffiltext{7}{INAF-Osservatorio Astronomico di Capodimonte, Salita Moiariello 16, 80131 Napoli, Italy}
\altaffiltext{8}{INAF-Osservatorio Astronomico di Bologna, Via Ranzani 1, 40127 Bologna, Italy}
\altaffiltext{9}{Department of Physics and Astronomy, Iowa State University, Ames, IA 50011, USA}
\altaffiltext{10}{The Observatories of the Carnegie Institution for Science, 813 Santa Barbara St., Pasadena, CA 91101, USA}
\altaffiltext{11}{INAF-Osservatorio Astronomico d'Abruzzo, Via Mentore Maggini snc, Loc. Collurania, 64100 Teramo, Italy}
\altaffiltext{12}{National Optical Astronomy Observatory, 950 N Cherry Avenue, Tucson, AZ 85719, USA}
\altaffiltext{13}{Department of Astronomy \& Astrophysics, University of Chicago, 5640 South Ellis Avenue, Chicago, IL 60637, USA}
\altaffiltext{14}{Max Planck Institute for Astronomy K\"onigstuhl 17 D-69117, Heidelberg, Germany}
\altaffiltext{15}{European Southern Observatory, Karl-Schwarzschild-Str. 2, 85748 Garching bei Munchen, Germany}
\altaffiltext{16}{Kiso Observatory, Institute of Astronomy, School of Science, The University of Tokyo, 
10762-30, Mitake, Kiso-machi, Kiso-gun, 3 Nagano 97-0101, Japan}
\altaffiltext{17}{Vatican Observatory, V00120 Vatican City State, Italy}
\altaffiltext{18}{Instituto de Astrof\'isica de Canarias, Calle Via Lactea s/n, E38205 La Laguna, Tenerife, Spain}
\altaffiltext{19}{INAF, Osservatorio Astronoico di Trieste, Via G.~B. Tiepolo 11, 34143 Trieste, Italy}
\altaffiltext{20}{INFN, Sezione di Pisa, Largo Pontecorvo 3, 56127, Pisa, Italy}
\altaffiltext{21}{Dipartimento di Fisica ``Enrico Fermi'', Universit\`a di Pisa, Largo Pontecorvo 3, 56127, Pisa, Italy}
\altaffiltext{22}{Stellingwerf Consulting, 11033 Mathis Mtn Rd SE, Huntsville, AL 35803, USA}
\altaffiltext{23}{Cerro Tololo Inter-American Observatory, National Optical Astronomy Observatory, Casilla 603, La Serena, Chile}
\altaffiltext{24}{Pontificia Universidad Catolica de Chile, Instituto de Astrofisica, Av. Vicu\~na Mackenna 4860, Santiago, Chile}

\date{\centering Submitted \today\ / Received / Accepted }

\begin{abstract}
We present a new complete Near-Infrared (NIR, $JHK_s$) census of RR Lyrae stars (RRLs)  
in the globular $\omega$ Cen (\ngc{5139}). We collected 15,472 $JHK_s$ images with 
4-8m class telescopes over 15 years (2000-2015) covering 
a sky area around the cluster center of 60$\times$34 arcmin$^2$.
These images provided calibrated photometry for 182 out of the 198 cluster RRL 
candidates with ten to sixty measurements per band. 
We also provide new homogeneous estimates of the photometric amplitude for 
180 ($J$), 176 ($H$) and 174 ($K_s$) RRLs.
These data were supplemented with single-epoch $JK_s$ magnitudes
from VHS and with single-epoch 
$H$ magnitudes from 2MASS. 
Using proprietary optical and NIR data together with 
new optical light curves (ASAS-SN) we also updated pulsation 
periods for 59 candidate RRLs. As a whole, we provide $JHK_s$ magnitudes 
for 90 RRab (fundamentals), 103 RRc (first overtones) 
and one RRd (mixed--mode pulsator). 
We found that NIR/optical photometric amplitude ratios increase when 
moving from first overtone to fundamental and to long-period 
(P$>$0.7 days) fundamental RRLs. 
Using predicted Period-Luminosity-Metallicity relations, we derive a true 
distance modulus of 13.674$\pm$0.008$\pm$0.038 mag (statistical error 
and standard deviation of the median)---based on spectroscopic 
iron abundances---and of 13.698$\pm$0.004$\pm$0.048 mag---based on photometric 
iron abundances. We also found evidence 
of possible systematics at the 5-10\% level in the zero-point of the PLs based on
the five calibrating RRLs whose parallaxes had been determined with \hst. 
\end{abstract}


\maketitle

\section{Introduction} \label{chapt_intro_omega}

There is mounting evidence that deep and accurate NIR photometry presents 
several indisputable advantages over optical photometry 
concerning distance determinations. The obvious
advantage is the lower sensitivity to reddening (i.e., uncertainties in the reddening values or the 
presence of differential reddening), but the advantages become even more relevant when dealing
 with primary distance indicators such as Cepheids (classical and type-II) and RR Lyrae (RRL) stars.  
Theory and observations indicate that the dispersions of Period-Luminosity 
(PL) relations steadily decrease when moving from optical to NIR bands. The PL 
relations can be derived neglecting the color term, i.e., the width in temperature 
of the instability strip, and this assumption becomes less severe in the NIR regime. 

Here, we will focus on the RRLs. In the optical ($BV$) bands they 
typically obey a magnitude vs metallicity relation, and the PL relation becomes 
evident only for wavelengths longer than the $R$ band \citep{marconi15,braga16}. 
The slope of the PL relation steadily increases from the $R$- to the 
$J$-band and attains an almost constant value for wavelengths longer than 2.2 
microns \citep{madore13,beaton2016,neeley17}.
The amplitudes display a similar trend: they attain their largest 
values in the $U$ band and approach an almost constant value for wavelengths 
longer than 2.2 microns \citep[][Marconi et al., in prep.]{braga15,neeley15}.
This empirical evidence and theoretical considerations both indicate that luminosity 
variation in the optical regime is mainly dominated by effective 
temperature variation, while in the NIR regime it is mainly 
dominated by radius variation \citep{madore13,bono2016a}.  
 
The metallicity dependence of RRLs, in contrast with classical 
Cepheids, is quite well established. Theory and observations 
indicate that an increase in metal content makes RRLs fainter.  
The above evidence makes RRLs key standard candles, and they provide  
a very promising opportunity to provide an independent calibration of 
secondary distance indicators and to 
constrain possible systematics between low-mass/old and 
intermediate-mass/young distance indicators \citep{beaton2016}. 
However, we still lack firm empirical estimates of the zero-point, 
the slope and the metallicity dependence of the diagnostics adopted 
to estimate individual RRL distances. 

In this context, cluster RRLs play a crucial role, since we have detailed 
knowledge of both the age and the chemical composition of their progenitors. 
In particular, the RRLs in $\omega$ Cen appear to be an ideal laboratory, 
even though we are still lacking solid constraints on the formation 
and evolution of this peculiar globular. The reasons are the following:

a) $\omega$ Cen includes almost 200 RRLs and they are almost equally split 
between fundamental and first overtone pulsators. This suggests that the 
instability strip is well populated both in the red/cool and in the blue/hot 
region.   

b) Current empirical evidence indicates that RRLs in $\omega$ Cen cover 
a metallicity range of at least one dex. This makes $\omega$ Cen
a fundamental testbed to constrain the metallicity dependence, since the 
depth effects are negligible compared with its distance. 

c) $\omega$ Cen hosts at least eight alternative distance indicators: 
1) tip of the red giant branch \citep{bellazzini04,bono2008b}; 
2) HB luminosity level \citep{vandenberg13}; 
3) Type II Cepheids \citep{matsunaga06,navarrete17}; 
4) Miras \citep{feast1965};
5) SX Phoenicis variables \citep{mcnamara00}; 
6) eclipsing binaries \citep{thompson2001}; 
7) the white dwarf cooling sequence \citep{ortolani_rosino1987,calamida2008}; 
8) an astrometric distance \citep{vandeven2006}. This provides a
unique opportunity to constrain the systematics affecting standard 
candles that originate from different physical mechanisms.  

In spite of the quoted indisputable advantages, the NIR investigations lag 
when compared with optical ones. Accurate NIR time series data for 
a significant fraction of $\omega$ Cen RRLs were provided for the 
first time by \citet{delprincipe06}. They adopted NIR time series data collected 
with SOFI at NTT and provided homogeneous mean $JK_s$ magnitudes 
for 180 variables (114 based on proprietary data: 81 {\it J},  
119 $K_s$ images).

More recently, \citet{navarrete15,navarrete17} collected NIR time 
series data (252 $J$ and 600 $K_s$ images) of $\omega$ Cen 
with the VIRCAM at ESO VISTA telescope. They discovered four new candidate 
RRLs (two cluster members and two nonmembers). They also provided new mean 
$JK_s$ magnitudes for 187 out of the 198 RRLs. 
Using NIR Period-Luminosity relations for both RRLs and Type II Cepheids 
they found a weighted-average true distance modulus to $\omega$ Cen of 
13.708$\pm$0.035 mag.

However, our datasets provide a few key advantages with respect to 
the quoted literature works: 1) full coverage of light curves 
in the $H$ band; 2) the possibility to complement our data with
proprietary optical data; 3) a better pixel scale that provides 
a more accurate photometric reduction of blended sources in the 
central region of the cluster.

Although the cluster and field RRLs have been at the crossroads of an
empirical effort of paramount importance (OGLE, CATALINA, Pan-STARRS, VVV) we still 
lack a detailed analysis of the pulsation properties (photometric amplitudes,  
topology of the instability strip) of RRL stars in the NIR regime. We are interested 
in providing a complete NIR census of RRLs in $\omega$ Cen as a stepping stone
for future developments.

i) To derive new and accurate NIR ($JHK_s$) template light curves. 
Future ground-based extremely large telescopes and space telescopes (JWST, 
EUCLID, WFIRST) will allow us to measure RRLs in Local Volume galaxies. 
It is plausible to assume that they will allow us to 
collect only a few random points, so NIR templates are essential to 
improve the accuracy of the mean magnitudes.   

ii) By taking advantage of the coupling between optical and NIR mean magnitudes, 
to provide a simultaneous estimate of distance, reddening, and metal 
content adopting an approach similar to that used by \citet{inno2016} for Classical
Cepheids in the Large Magellanic Cloud (LMC).

To accomplish these goals we took advantage of specific NIR ($JHK_s$) time 
series data collected with SOFI at NTT, with NEWFIRM at CTIO and with 
FourStar \citep{persson2013} at Magellan.

The structure of the paper is as follows. 
In \S~2 we present the NIR $JHK_s$  photometric data sets. In this section we introduce 
not only the NIR time series, but also show the cluster area covered by different datasets
and their NIR Color-Magnitude Diagrams (CMDs).  The entire sample of cluster RRLs is presented
in \S~3. Subsections 3.1 and 3.2 deal with the the phasing of the data (light curves) and 
with period estimates, while in subsection 3.3 we discuss the analytical fits to the 
light curves and the estimates of both mean magnitudes and photometric amplitudes. The comparison 
with mean magnitudes available in the literature is discussed in subsection 3.4.   
Subsection 4.1 introduces the topology of the instability strip both in NIR and in NIR-optical 
CMDs, while in subsection 4.2 we outline the properties of the variables in the luminosity 
amplitude versus logarithmic period (Bailey diagram) together with optical-NIR and 
NIR photometric amplitude ratios. In \S5 we present NIR PL relations and discuss in detail 
their dependence on the metal content.  The new distance determinations to $\omega$ Cen, 
based on NIR PL relations, and the comparison with literature estimates are discussed in \S~6.  
The summary of the main findings of the current investigation and the future developments 
of the overall project are outlined in \S~7. 

\section{NIR photometric data sets}\label{chapt_obs_nir}

A complete synopsis of our NIR datasets is given in 
Table~\ref{tbl:omegalog_nir}. The grand total of our images is 15,472:
5,102 in {\it J}, 4,872 in {\it H} and 5,498 in $K_s$, collected over 
15 years (January 2000 - January 2015).

The majority ($\sim$95\%) of our data was collected 
with the FourStar imager (pixel scale: 0.16 arcsec/pix) 
at the 6.5m Magellan-Baade 
telescope at Las Campanas during five nights in June 2013 
(10,800 images) and three nights in January 2015 (3,979 images; 
one exposure was missing the data from chip 3). 
The seeing during the 2013 run was better than 1.2\sec 
~90\% of the time and better than 0.85\sec ~half of the time.
Frames from the run of 2015 were collected in excellent
seeing conditions: 90\% of the times it was better than 0.6\sec ~and, 
half of the time the seeing was better than 0.45\sec. The 
fifteen pointings of the 2013 and 2015 data are almost the same, and cover
a sky area of 60$\times$34 arcmin$^2$ ($\sim$0.57 degree$^2$).
The dithering pattern is made up by five single exposures.

Around 5\% of our images were collected with the 
NEWFIRM instrument (0.4 arcsec/pix) at the CTIO 4.0m telescope
during one night in May 2010 (308 images) and with the SOFI camera (0.25 arcsec/pix)
at the NTT 3.6m telescope at La Silla (314 images) during 2000-2005. 
Only a few of these data (12 SOFI images) were collected in the $H$ band.
These data cover $\sim$1480 arcmin$^2$ (one NEWFIRM pointing and 
five SOFI pointings) and are completely contained within
the FourStar area. For 90\% of the time, the seeing was better than
1.0\sec, and half of the time better than 0.7\sec.
We point out that the images used by
\citet{delprincipe06} for their photometry are a subsample
(200 images) of our SOFI dataset.

Finally, we have also collected 70 images with MAD at 
VLT 8.0m, during 2007 April 3--5 and 2007 June 1 and 3.
These data were collected with a seeing of 0.7-0.9\sec but the 
AO unit of the instrument provided a mean FWHM smaller than 
0.2\sec for 90\% of the images and smaller than 0.1\sec for 
half of the images. The two pointings cover a sky area of 
2 arcmin$^2$.


As a whole, the entire NIR dataset for $\omega$ Cen exceeds the
capabilities of our computers for simultaneous reduction, so we performed DAOMASTER and 
ALLFRAME on four independent subsamples:
(1) the 2013 June Las Campanas data, hereinafter LCO13; 
(2) the 2015 January Las Campanas data, hereinafter LCO15; 
(3) the other natural-seeing observations, hereinafter ``other'';  
(4) AO assisted data, hereinafter MAD. 
After the completion of the profile-fitting photometry the four 
catalogs were merged to assign a common numbering scheme to the
individual stars.

The photometry was calibrated on the basis of 2MASS stars 
contained within our images.  We considered
only 2MASS All-Sky Point Source Catalog photometric 
measurements that had been assigned photometric
quality class ``A''.  This quality class corresponds 
to magnitude determinations in the $J$, $H$, or
$K_s$ bandpass with a signal-to-noise ratio$\,\geq\,$10 
($\sigma$(magnitude) estimated to be $\leq\,$0.109$\,$mag).
Matches between the 2MASS stars and entries in our joint 
catalog were determined by astrometric agreement
within a tolerance of 1~arcsecond; matches satisfying 
this criterion agreed positionally with a standard
deviation of 0.15 arcseconds in the $x$ ($\sim$ right ascension) 
direction and 0.14 arcseconds in $y$
($\sim$ declination). The 2MASS magnitudes were used as 
standard measurements to calibrate our instrumental
magnitudes using transformation equations employing linear 
color terms.  Individual stars displaying large
residuals from preliminary fits were gradually discarded until 
the transformation relied only on 12,802
individual 2MASS stars with fitting residuals $< 0.20\,$mag 
in at least one of the three bandpasses.

Our natural-seeing observations of $\omega$ Cen were calibrated 
to the 2MASS photometric system on the basis of
these transformation equations. An additional 172 
fainter but well-isolated stars that were well-observed
in our data sets were selected to serve as secondary calibrators for the MAD observations.

In reducing the NIR data set for $\omega$ Cen we adopted a double strategy. We 
performed ALLSTAR/ALLFRAME photometry over the entire set of NIR images. This 
approach is required to have accurate time series data to estimate the pulsation 
parameters.

Moreover, we also performed an independent photometric reduction based upon the stacked 
images. This approach was adopted to improve the detection of faint stars 
and to provide a very accurate and deep NIR ($JHK_s$) catalog, with no attempt at
time resolution.

\begin{figure*}[!htbp]
\centering
\includegraphics[width=11cm]{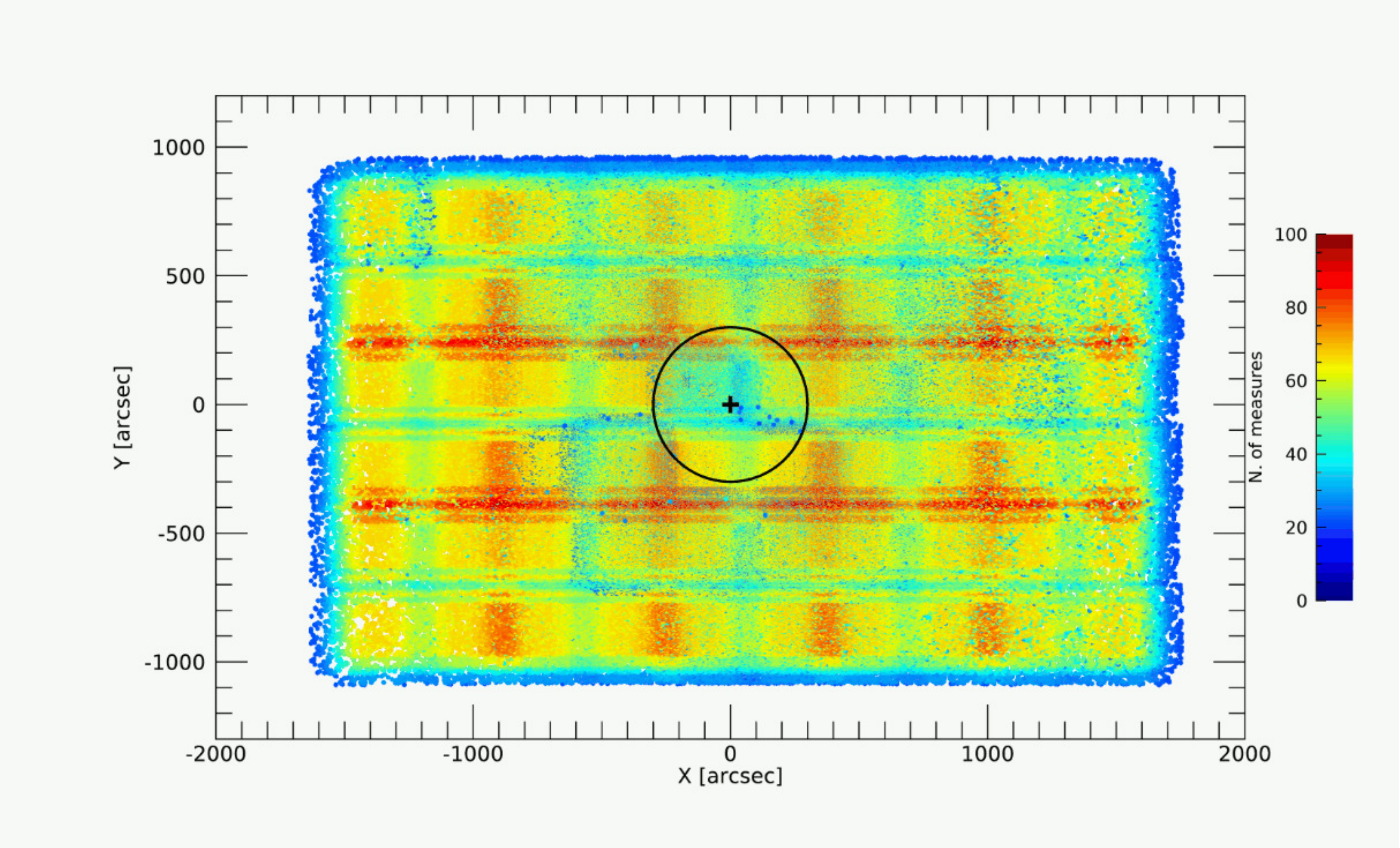}
\caption{
Distribution on the sky (arcseconds) of the photometry performed on
NIR ($JHK_s$) images collected with FourStar at Magellan in 2013 (LCO13). 
This data set covers a cluster area of 60$\times$34 arcmin$^2$. 
The color coding is correlated with the number of measurements 
(see the right bar). A black plus marks the position of the cluster center 
\citep{braga16}. The black circle marks the half-mass radius 
\citep[300 arcseconds][]{harris96}.
}
\label{fig:omegacen_13}
\end{figure*}

\begin{figure*}[!htbp]
\centering
\includegraphics[width=11cm]{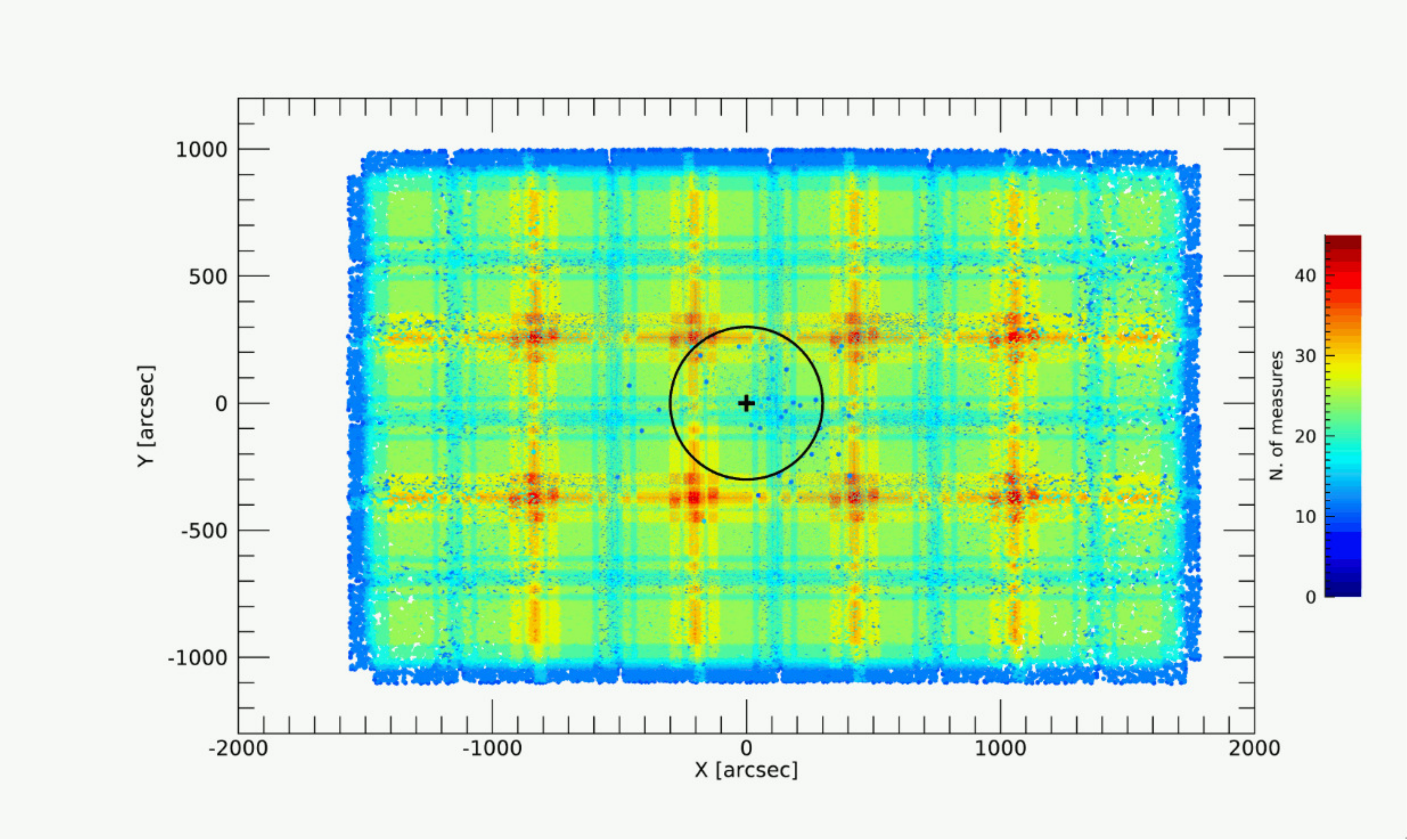}
\caption{ 
Same as in Fig.~\ref{fig:omegacen_13}, but for NIR ($JHK_s$) images collected 
with FourStar at Magellan in 2015 (LCO15). This data set covers a 
cluster area of 60$\times$34 arcmin$^2$.
}
\label{fig:omegacen_15}
\end{figure*}

\begin{figure*}[!htbp]
\centering
\includegraphics[width=11cm]{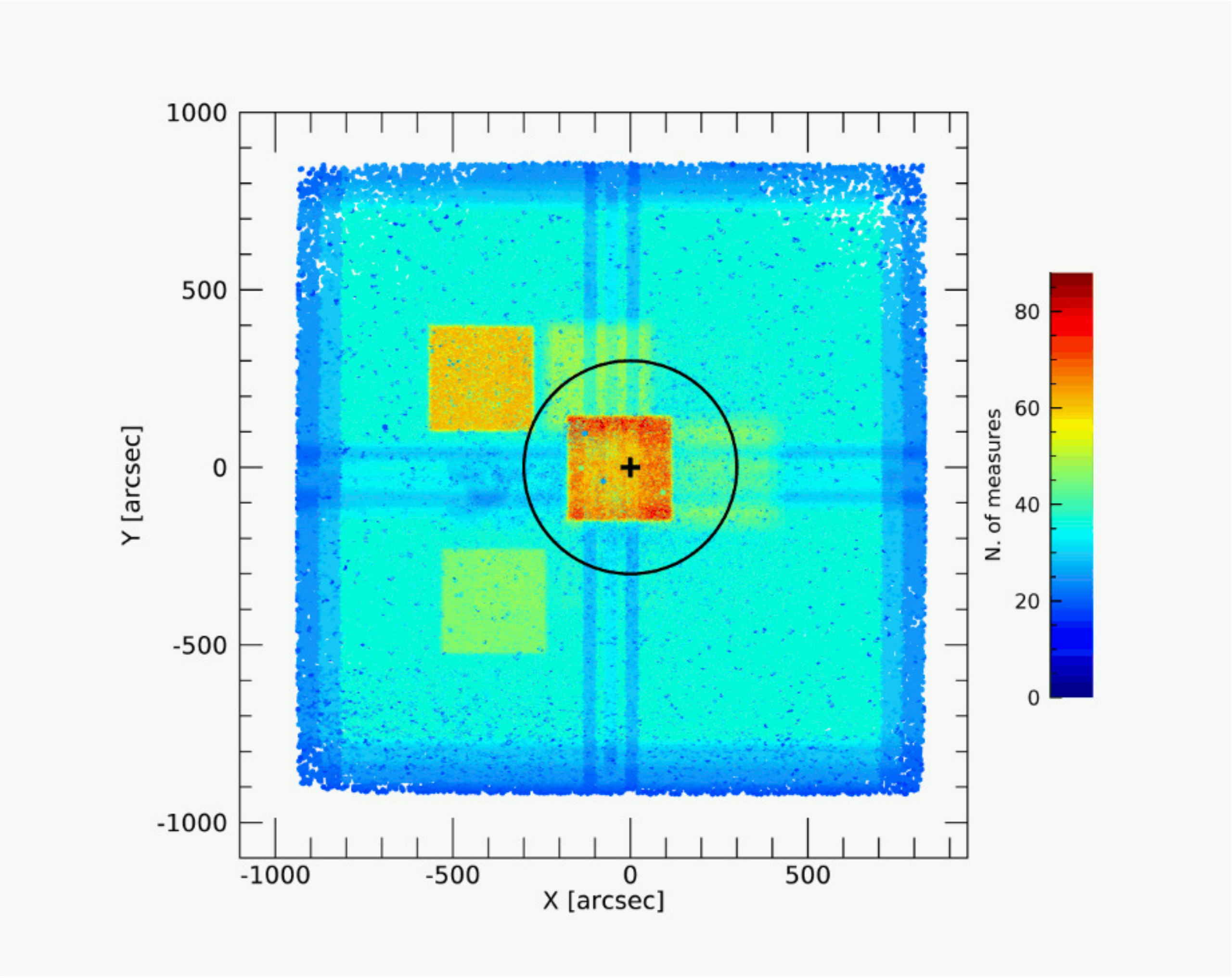}
\caption{ 
Same as in Fig.~\ref{fig:omegacen_13}, but for NIR ($JHK_s$) images 
collected with different telescopes (``other'' data set, see text for more details). 
This data set covers a cluster area of 27$\times$33 arcmin$^2$. 
}
\label{fig:omegacen_ot}
\end{figure*}
\clearpage

\begin{figure*}[!htbp]
\centering
\includegraphics[width=11cm]{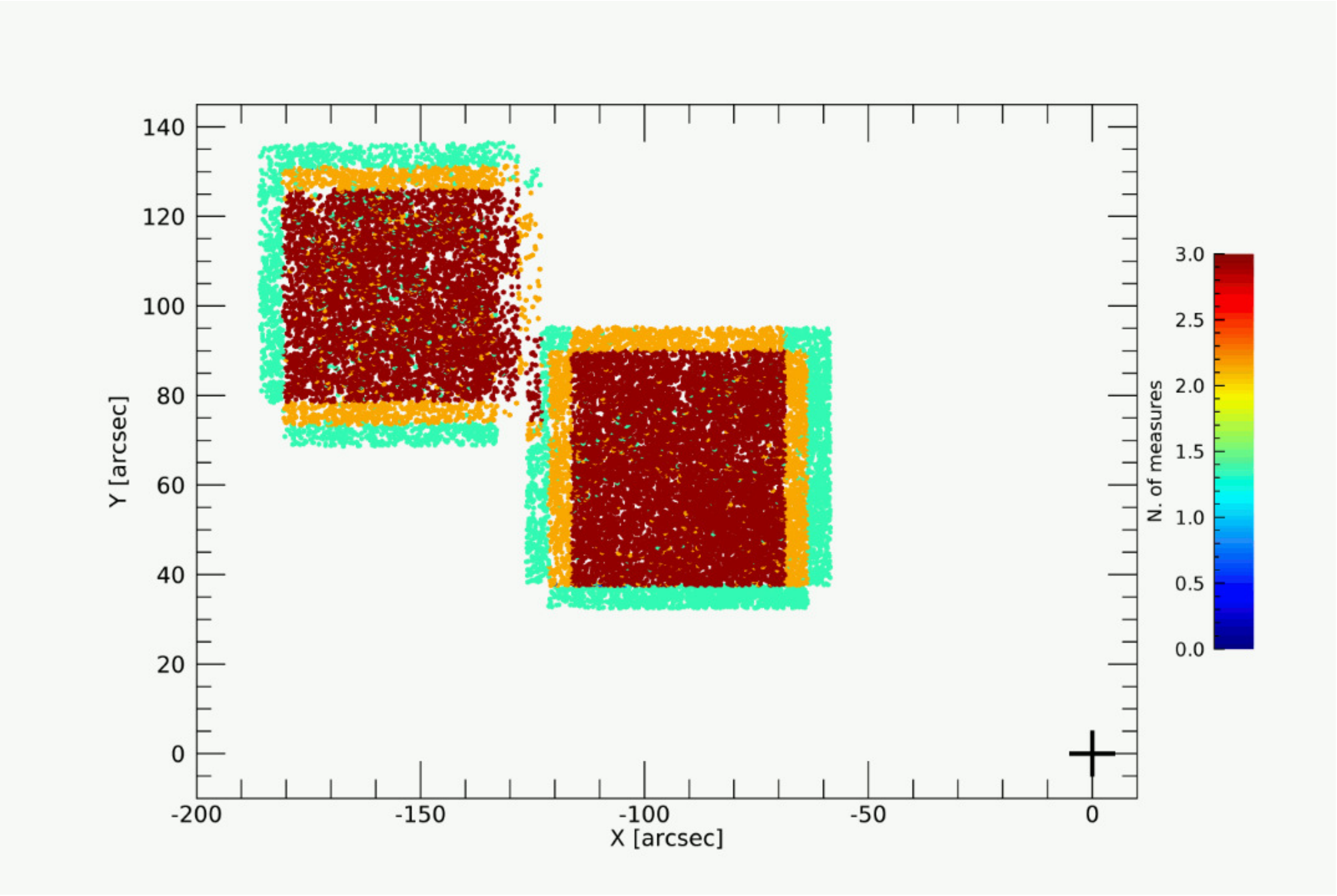}
\caption{ 
Same as in Fig.~\ref{fig:omegacen_13}, but for NIR ($JK_s$) images collected 
with the MCAO system (MAD) that was available at VLT. The individual 
pointings cover an area of 1$\times$1 arcmin$^2$.}
\label{fig:madmap}
\end{figure*}

\begin{figure*}[!htbp]
\centering
\includegraphics[width=16cm]{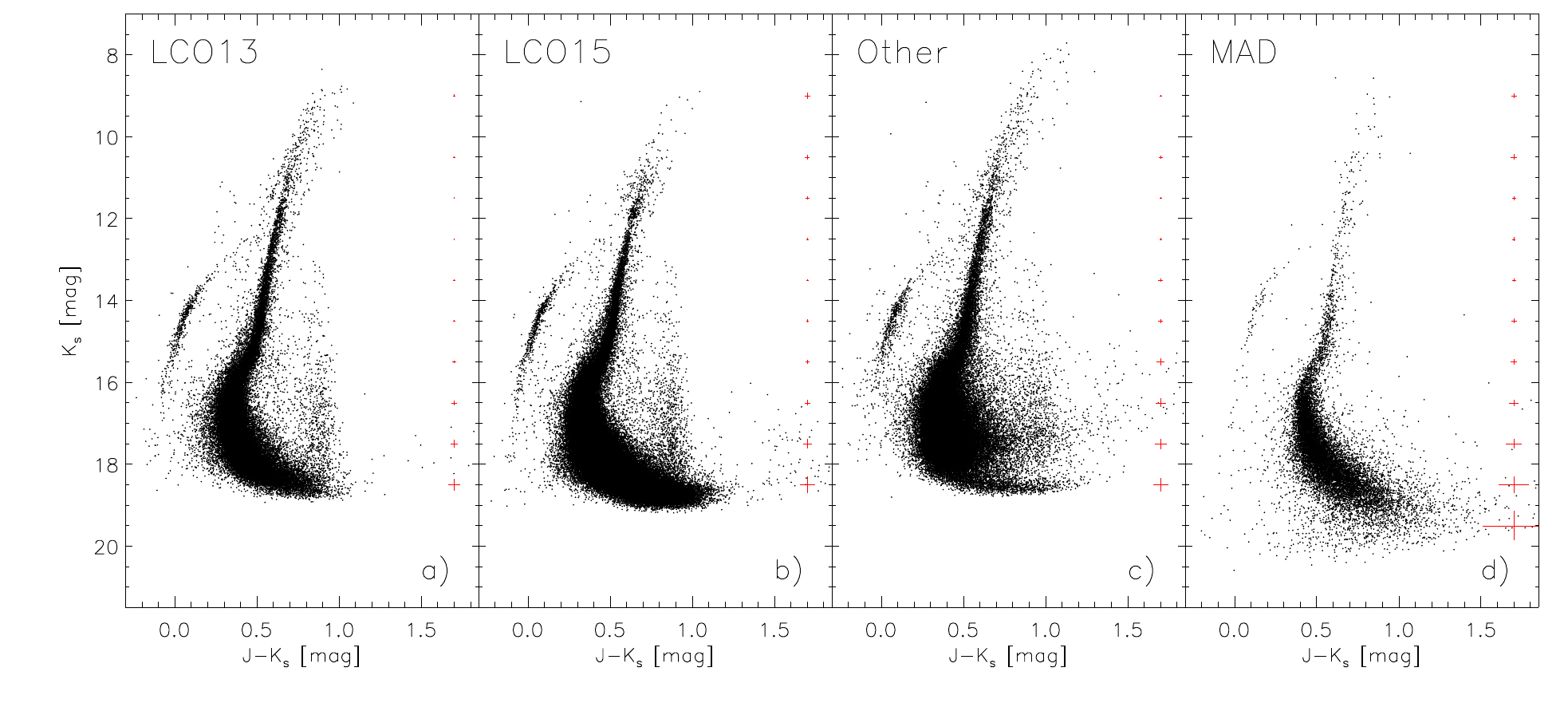}
\caption{NIR ($K_s$ vs $J-K_s$) color-magnitude diagrams of $\omega$ Cen.
Error bars display intrinsic errors both in magnitude and in color. To 
make them visible, they are magnified by a factor of five.}
\label{fig:omegacen_cmd_allcatalogs}
\end{figure*}

We have derived accurate CMDs from the current photometry,  
covering both the bright region typical of RGB and AGB stars 
(up to $K_s\sim$8.5 mag thanks to the SOFI data), 
but also $\sim$1.5-2.5 magnitudes 
fainter than the main sequence turn-off region
(FourStar data, see Fig.~\ref{fig:omegacen_cmd_allcatalogs}). 
The mean squared sums of the $J$- and $K_s$-band 
photometric errors in different magnitude bins 
are plotted as red error bars on the right of each panel.

The stars plotted in the above CMDs have been selected by 
the $\chi$ parameter---that quantifies the deviation between 
the star profile and the adopted Point Spread Function (PSF)---and 
the sharpness ($\bigl| sha \bigr|$ $<$ 0.7), that
indicates the difference in broadness of the individual stars compared 
with the PSF and is used to reject non-stellar sources. 
In passing we note that PSF photometry of 
individual images is essential to improve the precision of individual 
measurements of variable stars. The identification and 
fitting of faint sources located near the variable stars provides an 
optimal subtraction of light contamination from neighboring stars.

The effect of seeing is clear 
when we compare the CMDs in panels a) and b): 
the telescope and the camera are the same, but the better seeing
during 2015 allowed us to gain $\sim$0.3 mag in depth. 
However, the better seeing also causes 
a larger spread in color of stars on the upper RGB 
($K_s$<12 mag), due to a fainter level of saturation.

The effect of the seeing and of the pixel scale of the instrument
is also clear when comparing the LCO13, LCO15 and other datasets.
LCO15 is, in fact, $\sim$0.3 mag deeper than the other two and that 
with the least populated RGB and the sharpest
Blue and Extreme Horizontal Branch.
On the other hand, the other dataset has a better populated upper RGB.

Finally, the MAD dataset is the deepest, but most importantly,
shows the least amount of contamination by field stars---which 
is quite clear in the two LCO datasets at $J-K_s \sim$0.85 mag---since 
the observed sky area is small and very close to the cluster center. These 
are clear advantages of using AO corrections for the CMD observations
(but an obvious disadvantage for acquiring a large sample of RR Lyraes).


\section{RR Lyrae stars}\label{chapt_rrl}

We adopt Table 2 in \citet{braga16} as our reference list of RRL candidates
in $\omega$ Cen, but  exclude NV433 as suggested by \citet{navarrete15}. 
Note that we use the term ``RRL candidates'' because we lack 
solid, homogeneous constraints on their membership, other than 
considerations on their distance from the cluster center, mean
magnitude and proper motions for few of them \citep{vanleeuwen2000,bellini2009}.
Therefore, we kept 198 out of 199 objects from the quoted list.
According to the literature, eight of these 198 
stars are confirmed nonmembers \citep[V84, V168, 
V175, V181, V183, V283, NV457, 
NV458,][]{vanleeuwen2000,fernandeztrincado2015,navarrete15}. The classification of
V68 and V84 is ambiguous because, until now, their periods and their 
positions in both the CMD and in the Period-Magnitude plane 
did not allow a clear discrimination between RRL and Anomalous Cepheid classifications.
For three stars (V171, V178 and V179) neither periods
nor mean magnitudes were available, while for V182, only the period
was known. However, for the first time since these four stars were 
classified as variables \citep{wilkens1965,sawyerhogg1973}, we 
have retrieved multi-epoch photometry for them. Thus, we have 
data for all the RRL candidates of $\omega$ Cen.

For the master catalog for the ALLFRAME runs on our NIR data, 
we adopted the catalog generated during the photometric reduction of
the optical ($UBVRI$) data \citep[see \S2 ][]{braga16}. 
This allowed us to assign to the NIR point sources 
the same identification numbers as the optical sources, providing an unambiguous identification 
of all the stars within the sky area covered by our images and an accurate cross-match
of the optical and NIR catalogs. The cross-match allowed us to
retrieve the $UBVRI$ light curves of NV411, that we had missed 
in \citet{braga16}.

\subsection{Light curves}\label{par:lcvs}

The median number of phase points in the
light curves obtained from our NIR data is 109, 77 
and 136 for the $J$, $H$ and $K_s$ bands, respectively. 
The median of the photometric errors on the single-epoch magnitudes
are 0.012, 0.013 and 0.018 mag ($J$, $H$ and $K_s$). 
However, there is a large difference in the photometric errors 
between the FourStar data (median $\ltsim$0.015 mag in the $JHK_s$ bands) 
and the datasets $blanco$ and $milena$ (subsets of 
the other dataset, Table~\ref{tbl:omegalog_nir}, median 
$\gtsim$0.025 mag in all bands); moreover, errors on the 
single phase points can be as large as 0.2 mag.

As a preliminary step before analyzing the light curves, we binned the phase 
points to combine data from the same dithering sequence. The time step was 
$\sim$90 seconds for the LCO dithering sequence and $\sim$900 seconds 
for the other data.  We tested different binning methods, including simple 
intensity mean, median and weighted intensity mean. We adopted the weighted 
intensity mean because it provides smoother light curves. 
Note that, before averaging the single phase points, we performed a 
sigma clipping at a 3$\sigma$ level to reject the outliers. The rejected phase 
points were probably obtained from images with either a lower quality or for 
which the photometric solution was not optimal. The fraction of phase points 
that was rejected in this step is smaller than 5\%.

Note that the binning of the data was applied to the dithering sequences 
of individual data sets. The duration of a dithering sequence is a tiny 
fraction of the pulsation period of an RRL. Moreover, the bulk of the 
NIR data were collected as time series data, therefore, possible changes 
in the pulsation period minimally affect the binning of the data.  
After the binning process, we ended up with light curves including a 
median number of phase points of 25 ($J$), 17 ($H$) and 30 ($K_s$), 
respectively.

\subsection{Pulsation periods}\label{par:period}

We have already mentioned in Section~\ref{chapt_obs_nir} that the majority of 
our NIR data were collected in 2013 and in 2015. On the other hand, most of the 
optical time-series data were collected between 1995 and 1999. The 
remarkable overall time coverage (27 years) of optical plus NIR data 
allowed us to revise the period estimates derived in \citet{braga16}.
We have derived the new period estimates by adopting
the same method as the quoted paper, 
based on the Lomb-Scargle periodogram \citep{scargle82}.
This method simultaneously folds the time series of 
all the photometric bands that are available. A visual 
inspection of the folded light curves and a quantitative
estimate of the $\chi^2$ allow us to estimate the period.
On the basis of our overall optical and NIR photometry,  
we provide new period estimates for 59 variables. 
For 17 of them, the new period estimate should 
be considered as an improvement of 
the estimate based only on the optical data. 
For 30 variables, the new estimate suggests an intrinsic period change 
occurred during the time that passed between the 
acquisition of the bulk of the optical and NIR data 
(as indicated in Table~\ref{tab:phot_rr_nir}).
For the remaining twelve variables, the period available in 
the literature was updated. Note that V171 and V179 had no period 
at all in the literature.

We point out that our ability to detect period variations 
for some variables is a consequence of the accuracy of 
our method to estimate periods. A full analysis of 
period variations is not within the scope of this paper, so
we discuss only a few variables. 
An extensive analysis of the rate of period change ($\beta$) of the 
RRLs in $\omega$ Cen was performed by \citet{jurcsik2001}.
Among the seven variables (V68, V101, V104, V123, V150, V151 and V160)
for which \citet{jurcsik2001} detected a high $\beta$ 
(> 50$\cdot$10$^{-10}$ days/day), five (V68, V104, V123 and V150) 
do show a difference in our new and old period estimates. 
We do not have enough NIR data to detect period variations 
for V151 and V160.
One extreme case is V104, since we observe a period variation as 
large as $\sim+$0.001 days in only $\sim$15 years. 
According to both \citet{jurcsik2001} and this work, this is the RRL with
the highest $\beta$ in $\omega$ Cen.

We have also derived new periods for 
13 out of the 14 candidate RRLs that lie 
in the outskirts of the cluster and are not covered by our images. 
To estimate their pulsation periods, we took advantage of the $V$-band 
data collected by the ASAS-SN \citep{shappee2014,kochanek2017} survey
(\url{https://asas-sn.osu.edu/}). 
We have obtained a period estimate for all targets but V178, which shows 
no variability.  We point out that, for the first time, we have
derived the period of V171 and V179: these were
classified as RRa and RRb respectively by 
\citet{wilkens1965}, but the periods were not
published. We confirm that V171 is a RRab star and that
it is a cluster member. V179, however, is an eclipsing
binary. Finally, we have corrected the coordinates 
of V182---for which no finding chart exists---provided 
by \citet{Clement01} on the basis of the \citet{sawyerhogg1973}
catalog. We confirm that V182 is a RRab star but is not 
member of the cluster. For details, see Appendix~\ref{individualnotes}.

\subsection{Mean magnitudes and amplitudes}\label{par:lcvs}

In the literature, the sampling of NIR light curves of variable 
stars is far from being ideal. This is the main reason why template light curves have been 
developed for both RRLs \citep{jones1996} and classical Cepheids 
\citep{soszynski05,inno15}. The phase coverage of the 
current data set is quite good and ranges from around ten to 
sixty measurements per band. However, 
for variables located in the outskirts of the cluster (four RRLs between
20\min ~and 32\min ~from the center of the cluster) we 
have less than ten measurements. For an additional 16 RRL candidates farther away,
we only have retrieved one measurement from both  
VHS \citep[$JK_s$,][]{mcmahon13} and 2MASS \citep[$H$,][]{skrutskie2006}. 
In order to quantify the impact that analytical fits of the light 
curves have on the mean magnitudes and on the photometric amplitudes 
we decided to use three different approaches.  

{\em Locally weighted polynomial regression, PLOESS}-- A similar
method (GLOESS) has already been applied to fit randomly sampled light curves 
\citep{persson2004,neeley15,monson2017}. The key idea of this approach 
is to provide a plausible guess of the fitting function in a phase range 
for which the sampling is either too coarse or too noisy to allow fitting a low-degree 
polynomial to a subset of the data. Moreover, to limit the contribution 
of possible outliers, the individual points are weighted with a weight function.  
This is the reason why PLOESS is a local weighted regression method. 

The algorithm we developed relies on the following steps.
Let us assume that the time series data of the variable we are dealing 
with consists of $x_i$ (phase), $y_i$ (magnitude) data with i=1,\ldots,n 
phase points. The original data are divided into 
subsamples, each one including $\approx$20-30\% of the entire data set. 
Moreover, the weights for the individual data points in the subsample 
are defined using the following formula:\\  

$W_i = ( 1 - abs([X - x_i] / \Delta X)^3 )^3$\\

where $X$ is the phase at which we would like to have a new smoothed value 
along the light curve, $x_i$ are the data in the subsample and $\Delta X$ 
is the maximum distance in phase between $X$ and the data in the subsample. 
The weights were defined in such a way that the data point to be smoothed 
(X) has the largest weight. The weights ($W_i$) of the data points in the subsample 
($x_i$) decrease as a function of their distance from $X$, while the data points 
not included in the subsample have zero weight. The weighted least-squares 
regression on the data of the subsample is performed using a second 
degree polynomial and provides a new value of the light curve at the 
phase of the data point ($X$) we are smoothing. To overcome the typical problems 
at the boundaries of the phase interval [0,1], the data points were triplicated, 
i.e., the smoothing was performed on data replicated over the phase interval [-1,2]. Moreover, the 
data points included in each subsample are symmetric, i.e., the number of data 
points to the left and to the right of the data point (X) is always the same.   
We performed a number of simulations using also the weight function suggested 
by \citet{cleveland1979}, but we found that the current weight function provides 
smoother light curves when the original data points are grouped in restricted 
phase intervals. Finally, to further improve the stability of the fit, we also 
computed the residuals of the original data points from the smoothed light curve 
and, using an iterative procedure, we neglected from the smoothing the data points 
that are located at a distance larger than six times the median absolute 
value\footnote{This algorithm was implemented in IDL and it is available 
upon request to the authors.}.

{\em Spline}-- This is a classical approach with the key advantage of 
tightly fitting the data points. However, this approach is more prone to 
systematic errors when the time series data are either unevenly 
sampled or characterized by significantly different random errors. 
The spline fits, too, were derived on a triplicated light curve 
with phases in the interval [0,3] to avoid boundary effects. We 
adopted the middle section---phases in the interval [1,2]---of the spline 
fit as the final fit of the light curve.

The two quoted approaches were adopted to fit $J$-, $H$- and $K_s$-band
light curves. The PLOESS approach was adopted only for light curves
with a number of phase points larger than nine.

\begin{figure*}[!htbp]
\centering
\includegraphics[width=14cm]{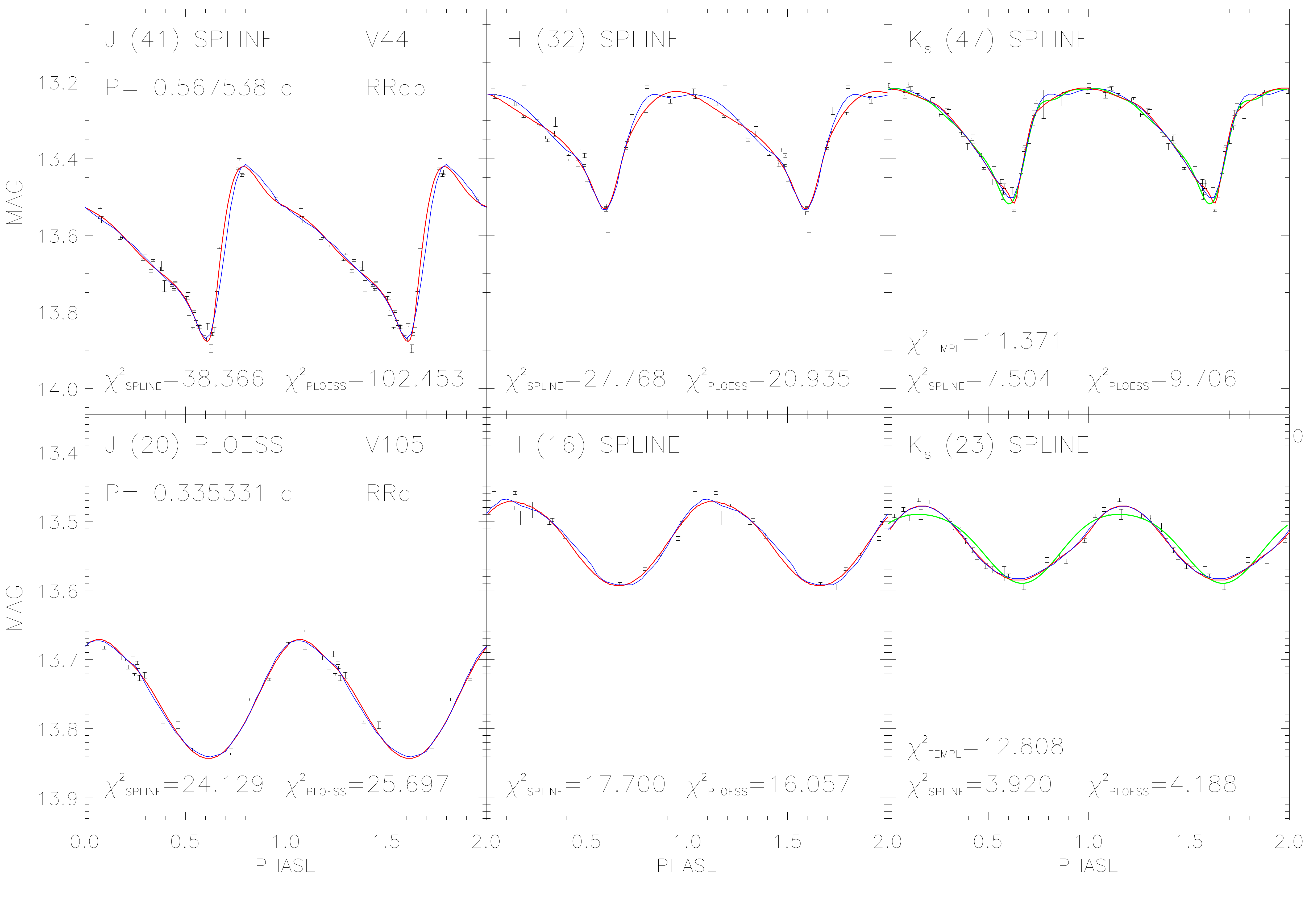}
\caption{Top: Light curve for the RRab variable V44. The red line
shows the spline fit, while the blue line the PLOESS fit 
and the green line the template fit. The vertical error bars display the 
intrinsic photometric error. The name and the period of the 
variable are labelled in the top-left panel. In the top-left corner
of each panel, the fitting model (SPLINE, PLOESS) that was selected 
as the best one, is labelled.
Bottom: Same as top, but for the RRc variable V105.}
\label{fig:lcv1}
\end{figure*}

\begin{figure*}[!htbp]
\centering
\includegraphics[width=14cm]{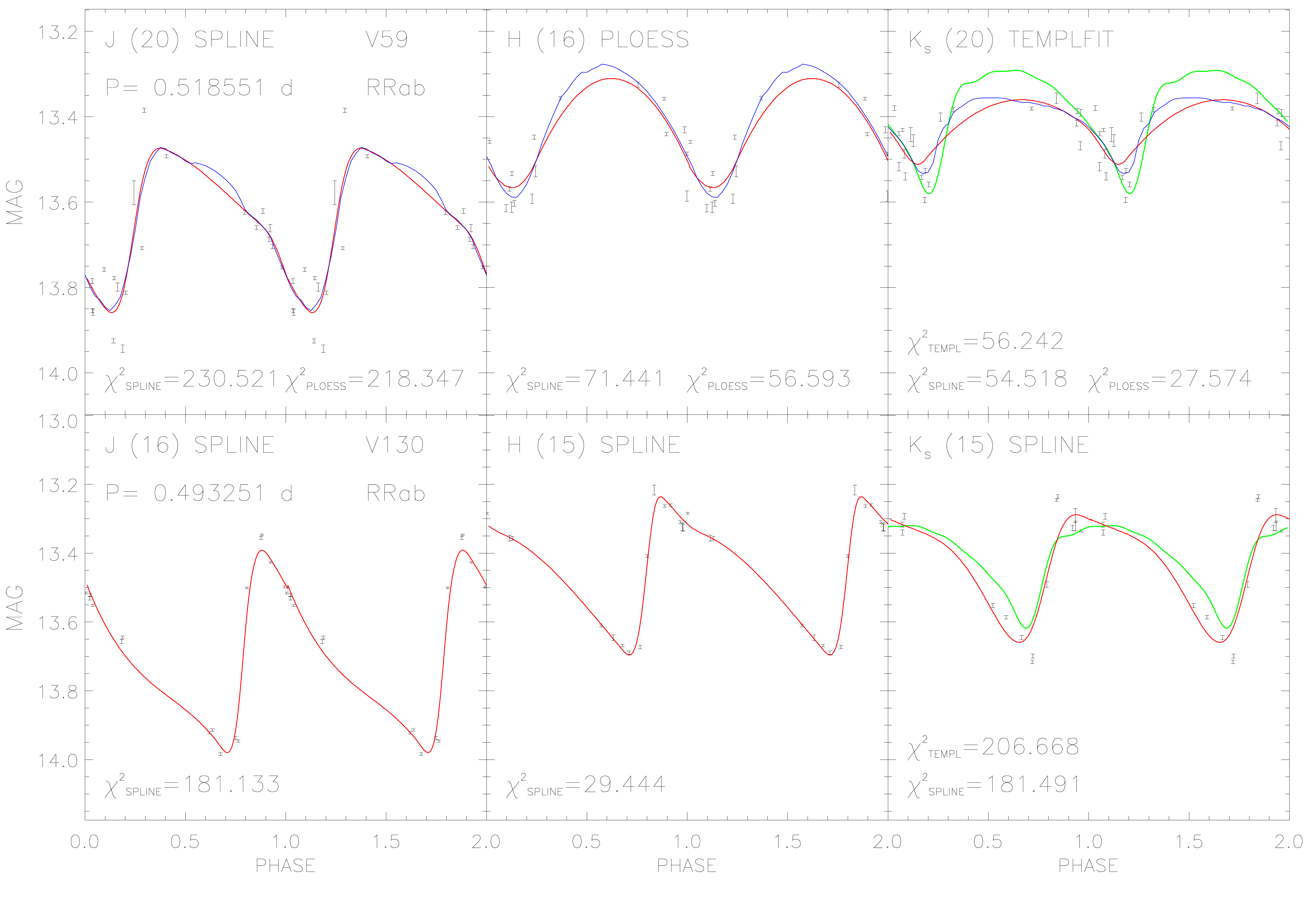}
\caption{Top: Same as top in Fig.~\ref{fig:lcv1}, but for the 
RRab variable V59. 
Bottom: Same as top, but for the RRab variable V130.}
\label{fig:lcv2}
\end{figure*}

{\em Template}-- The $K_s$-band light curves were also fit with the template light curves 
provided by \citet{jones1996}. We have found that for more than 60\% of 
the light curves in our sample, the mean magnitudes based on the template 
fit are, within the uncertainties, very similar to those based on the 
Spline and on the PLOESS fit.  However, this method is extremely sensitive 
to the accuracy of the period, to possible period variations, and to phase 
modulations (mixed-mode, Blazhko). 
Indeed, for more than $\sim$35\% of RRL candidates we found a phase 
shift between the template light curve and the observed data points. To 
overcome this limitation  we have adopted a different approach to apply 
the template fit. The first two steps are the same as in
\citet{jones1996}: first, we selected the template based
on the pulsation mode and on the optical $AB$ (or $AV$) amplitude; 
second, we set the scaling factor of the template fit as half of 
the $K_s$-band amplitude, calculated as $AK_s=0.108\cdot AB + 0.168$ 
mag (RRab) and $AK_s=0.110$ (RRc). 
The third step---the phasing of the template---is different: 
instead of anchoring the template to
one of the phase points---as in the canonical template fit---we 
have minimized the residuals ($\chi^2$) using two free parameters:
the mean magnitude and the phase shift. 
Note that to further improve the accuracy of the 
fit we could have used the $AK_s$ amplitudes evaluated using either 
the Spline or the PLOESS fit. We followed the classical approach to 
test both the accuracy and the precision of the template light curves. The 
above findings call for the development of new template 
light curves, and in particular for extension of the template 
light curves to the $J$ and $H$ bands.
Figure~\ref{fig:lcv1} shows the light curves in the $JHK_s$ 
bands of a RRc and a RRab variable with good phase coverage. Spline, 
PLOESS and template fits are also displayed. On the other hand, 
Figure~\ref{fig:lcv2} shows the light curves in the $JHK_s$ 
bands of two RRab variables either with a modest number of phase points or 
with gaps in the phase coverage. The $K_s$-band light curve of V59 is best 
fitted by the template, while for V130 the best fit is given by the spline.

Once the fits to the light curves were performed (Spline, PLOESS, 
template), we derived the mean magnitudes, photometric amplitudes, 
and their uncertainties. Note that the mean magnitudes were
derived by converting magnitudes to intensity, in arbitrary units,
then averaging and re-converting to magnitudes. 
For the RRLs for which either the phase coverage is not optimal or the light 
curve is too noisy, the final value of the mean magnitude was estimated as 
the median of the magnitudes converted to intensity over the individual phase 
points.
Photometric amplitudes were derived by the difference 
between the maximum and minimum of the fit of the light curve, whereas  
the mean magnitude was derived by either the Spline or the PLOESS fit.
The uncertainty on the photometric amplitudes were estimated summing in 
quadrature the median photometric error of the phase points around the minimum 
and the maximum of the light curve plus the standard deviation of the same 
phase points around the fit of the light curve. The final value was 
weighted with the number of phase points around minimum and maximum phases. 
The error on the amplitudes, for the few light curves with no phase points 
either across minimum or maximum phase, were estimated by summing in quadrature 
the difference between the maximum/minimum of the fit and the faintest/brightest 
phase point.
Note that we do not provide the photometric amplitude
from template fit, since this parameter is an input in
this approach. The mean magnitudes, the amplitudes, and their errors 
from the fits of the light curves 
are listed in Table~\ref{tab:phot_rr_nir}. 
Figures~\ref{fig:lcv1} and \ref{fig:lcv2} 
also display a pseudo-$\chi^2$ statistics for each fit:\\

$\chi^2 = \dfrac{1}{N} \sum_{i=1}^{N} \dfrac{( mag(obs)_{i} - mag(fit)_{i} )^2}{err(obs)^2_{i}} $\\

Taking account of the smallness of the photometric errors 
($err_{obs}$), the derived $\chi^2$ are low 
(see Figs.~\ref{fig:lcv1} and \ref{fig:lcv2}). This is further
evidence of the goodness of fits of the light 
curves. Note that, on the basis of both
a visual inspection of the light curves, 
and the comparison of the uncertainties 
on the means magnitude and amplitudes, we selected for each RRL the 
most accurate fit among Spline, PLOESS and template. Note that
the fits that we selected are not always those with the smallest $\chi^2$.

\begin{figure*}[!htbp]
\centering
\includegraphics[width=11cm]{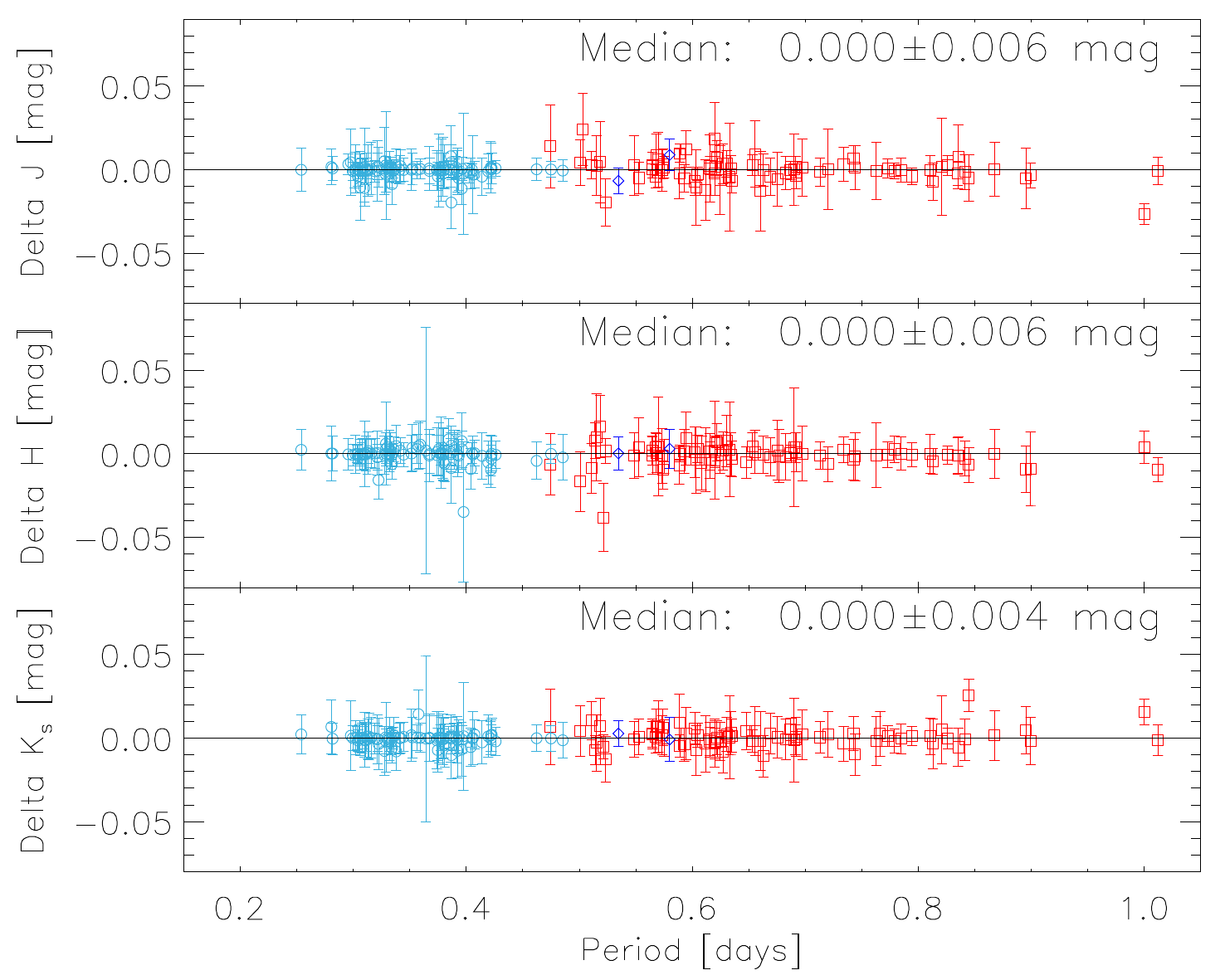}
\caption{Top -- Difference between the mean $J$-band magnitudes estimated  using 
the Spline and the PLOESS fit as a function of the pulsation period.
Light blue circles and red squares mark 
RRc (first overtone) and RRab (fundamental) variables. 
Dark blue diamonds mark variables of uncertain type.
Middle -- Same as the top, but for the $H$-band. 
Bottom -- Same as the top, but for the $K_s$-band. 
}
\label{fig:delta_spline_poly}
\end{figure*}

\begin{figure*}[!htbp]
\centering
\includegraphics[width=11cm]{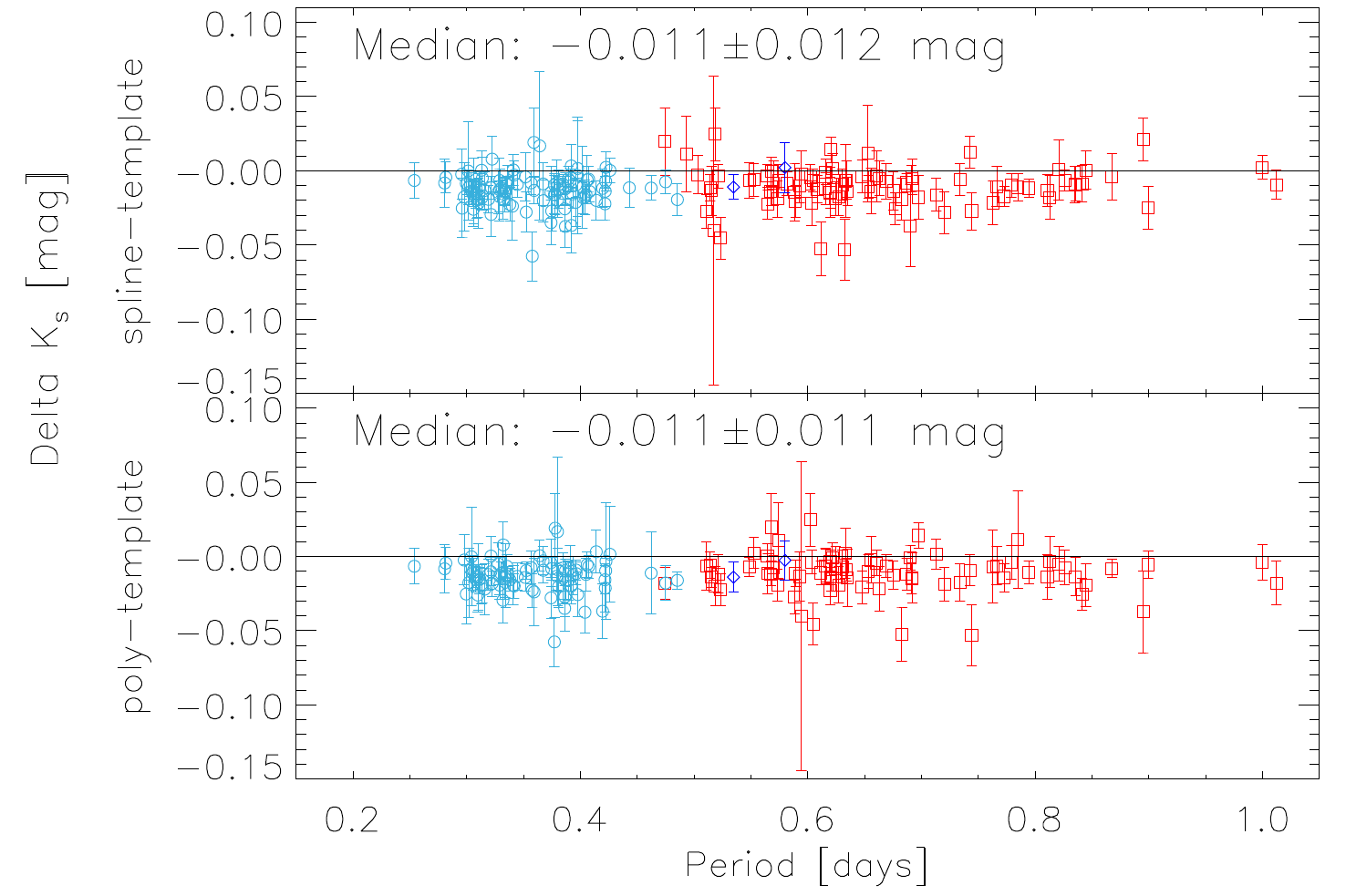}
\caption{Top -- Difference between the mean $K_s$-band magnitudes derived 
using the Spline and the template fit as a function of the pulsation period. 
Symbols are the same as in Fig.~\ref{fig:delta_spline_poly}.
Bottom -- Same as the top, but the difference is between the  
PLOESS and the template fit. 
}
\label{fig:delta_template}
\end{figure*}

Figures~\ref{fig:delta_spline_poly} and \ref{fig:delta_template}
show the difference in mean magnitude between the three different 
methods adopted to fit the light curves.
Spline and PLOESS provide mean magnitudes that are 
very similar, indeed the difference and the standard deviations 
are vanishing ($\Delta J$=0.000$\pm$0.006, $\Delta H$=0.000$\pm$0.006, 
$\Delta K_s$=0.000$\pm$0.004 mag). However, photometric amplitudes 
display more significant differences between Spline and PLOESS fits: 
they range from a few thousandths for low-amplitude 
RRc variables to one or two tenths of a magnitude for short-period, large amplitude 
RRab variables. The difference for the latter group is mainly caused by the fact that 
the PLOESS fits better represent the ripple across the minimum light phases than the Spline.  

We note that template fits have at least three disadvantages. One is the 
aforementioned sensitivity to the period: an error of 10$^{-5}$ days or 
larger may lead to a wrong phasing of the light curve, even
with accurate epochs of maximum. Moreover the light curve 
template of RRc stars has a fixed amplitude because it is based
only on four of them and it was not possible to establish a 
relation between optical and NIR amplitudes. Finally, 
as shown in Fig.~\ref{fig:delta_template}, the mean magnitudes 
obtained from the template fits are $\sim$0.01 mag fainter than 
those obtained with Spline or PLOESS fitting. Most
of the times, this difference is caused by the deeper minima 
of the template fits with respect to the other fitting functions.

\subsection{Comparison of NIR mean magnitudes}\label{par:magcompare}

\begin{figure*}[!htbp]
\centering
\includegraphics[width=11cm]{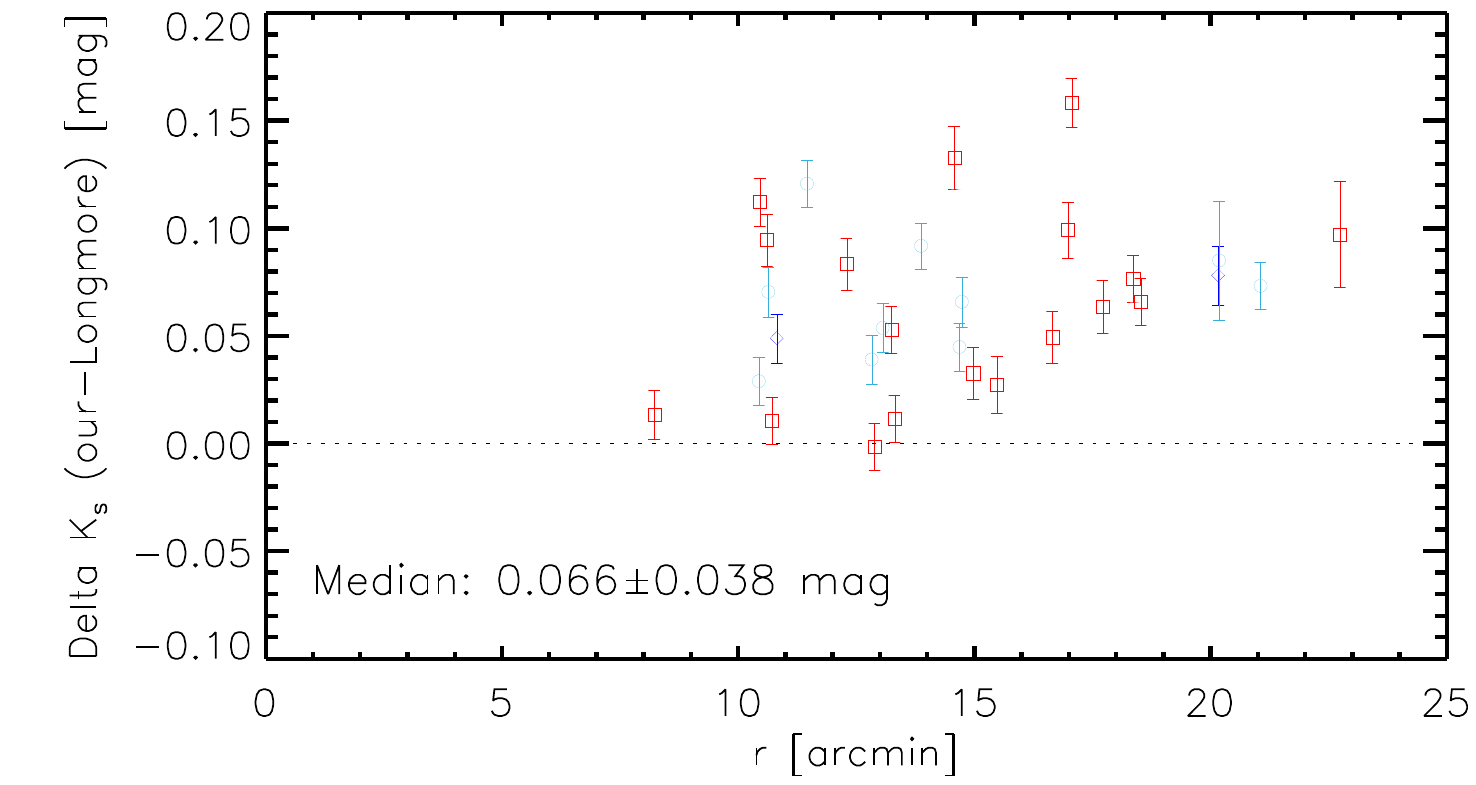}
\caption{Difference between our final $K_s$-band mean magnitudes 
and those derived by \citet{longmore1990}
as a function of the radial distance from the center of the cluster. 
The label depicts the median and the standard deviation of the 
sample. The vertical bars display the sum in quadrature 
of the uncertainties on mean magnitudes of the two datasets.
Symbols are the same as in Fig.~\ref{fig:delta_spline_poly}.
}
\label{fig:delta_longmore}
\end{figure*}

\begin{figure*}[!htbp]
\centering
\includegraphics[width=11cm]{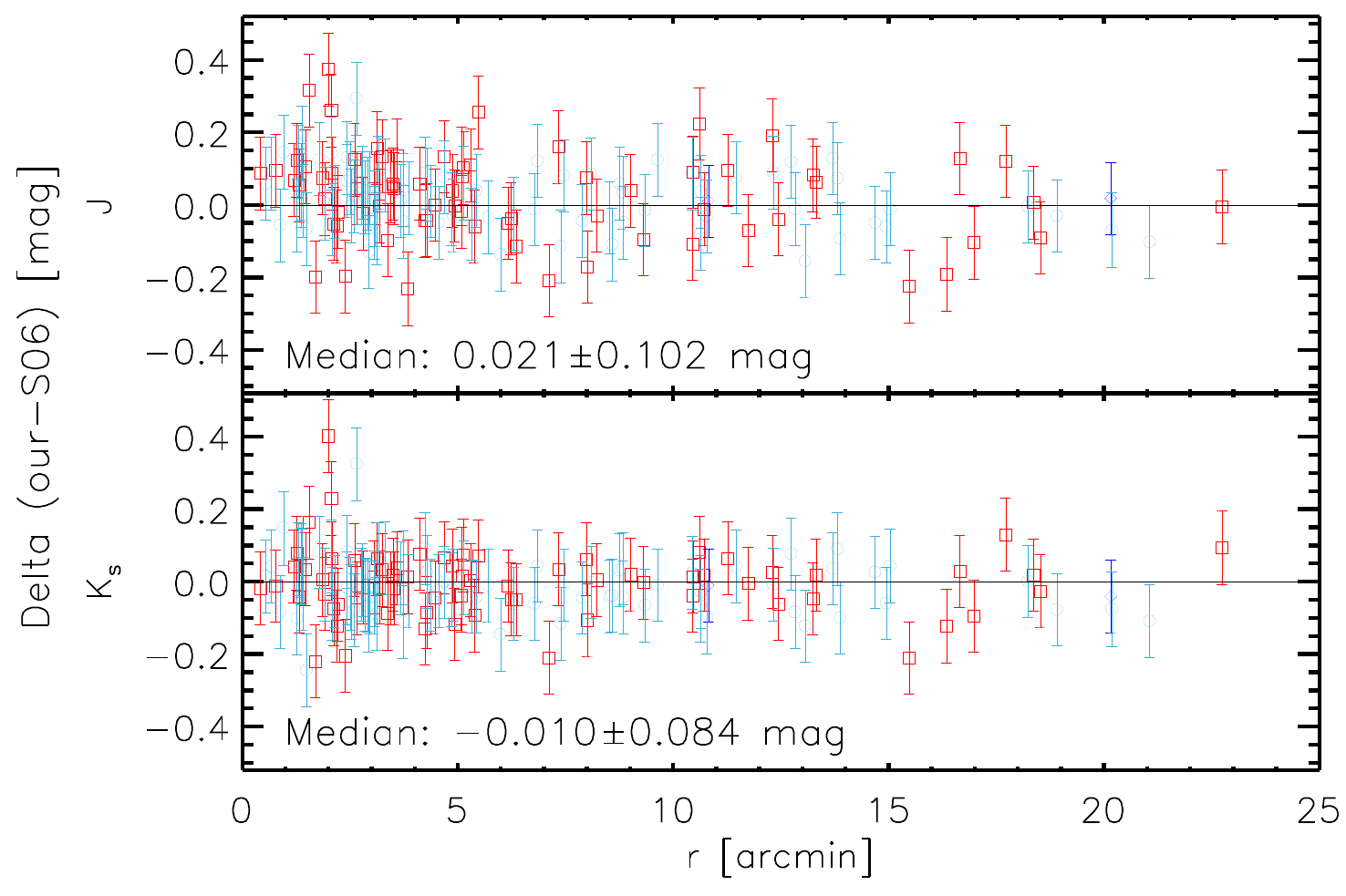}
\caption{Top -- Difference between our final $J$-band mean magnitudes 
and those derived by \citet{sollima06a} as a function of the radial 
distance from the center of the cluster. The label depicts the median 
and the standard deviation of the sample. The vertical 
bars display the sum in quadrature 
of the uncertainties on mean magnitudes of the two datasets
Symbols are the same as in Fig.~\ref{fig:delta_spline_poly}.
Bottom -- Same as the top, but for the $K_s$-band mean magnitudes.}
\label{fig:delta_sollima}
\end{figure*}

\begin{figure*}[!htbp]
\centering
\includegraphics[width=11cm]{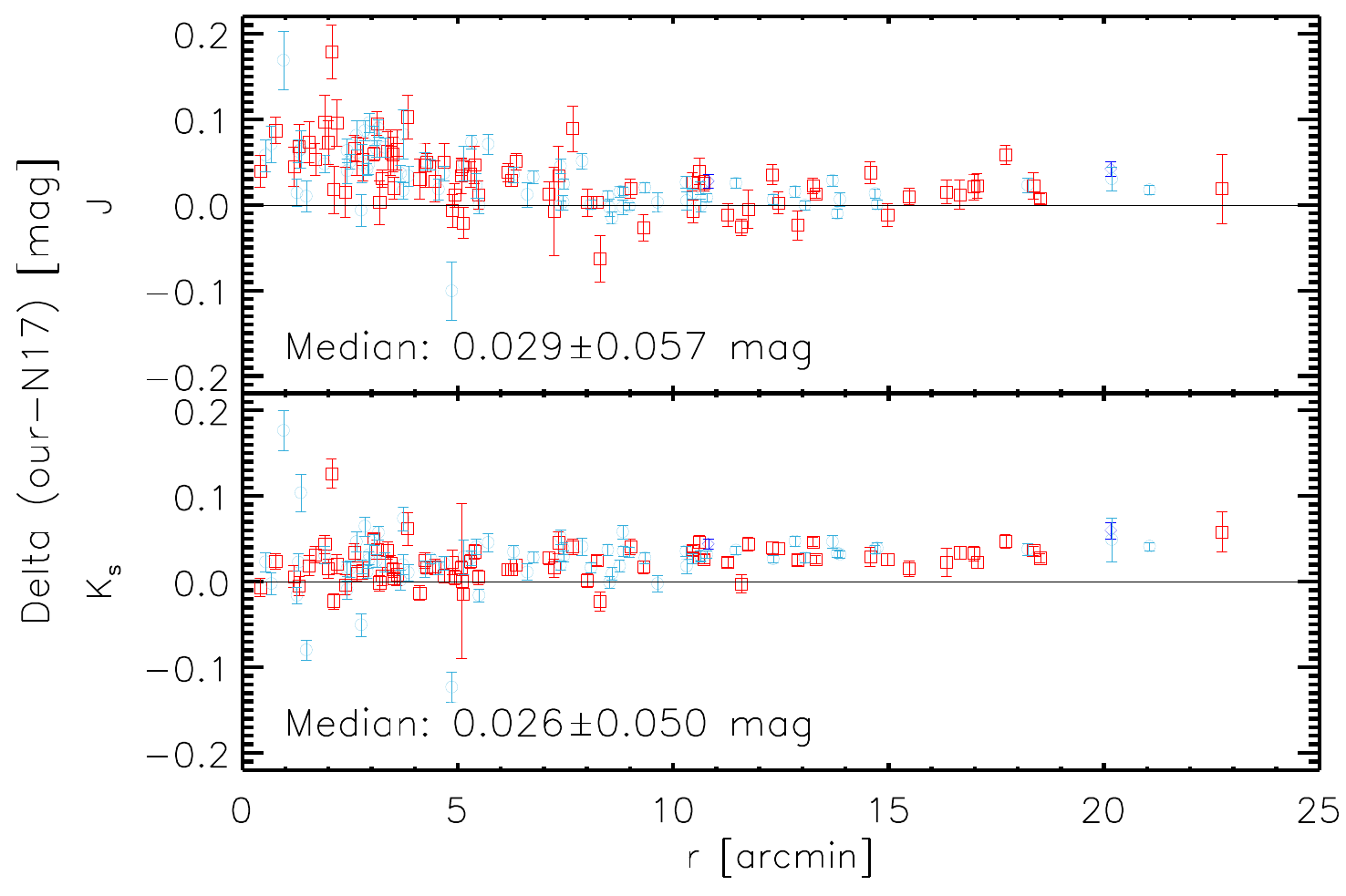}
\caption{Top -- Difference between our final $J$-band mean magnitudes 
and those derived by \citet{navarrete17} as a function of the radial 
distance from the center of the cluster. The label depicts the median 
and the standard deviation of the sample. The vertical 
bars display the sum in quadrature 
of the uncertainties on mean magnitudes of the two datasets
Symbols are the same as in Fig.~\ref{fig:delta_spline_poly}.
Bottom -- Same as the top, but for the $K_s$-band mean magnitudes.}
\label{fig:delta_nava}
\end{figure*}

We compare the mean magnitudes that we obtain with those
of \citet{longmore1990, sollima06a,navarrete17}. To do the 
comparison in the same photometric system (2MASS), we have adopted
the transformations of \citet{carpenter2001} to convert 
the AAO mean magnitudes of \citet{longmore1990}. Since these 
transformations need $J-K_{AAO}$ as an input, but the $J$-band magnitude
is not provided by \citet{longmore1990}, we have adopted 
$J-K_{AAO}$=0.25 mag for all RRLs. This is an approximate mean
of the $J-K_s$ colors of RRLs. We point out that a shift of 
0.05 mag in $J-K_{AAO}$ means a change of 0.01 mag in the 
output $K_s$-band magnitude. The $J-K_s$ colors of 
RRLs range from 0.15 and 0.40 mag, therefore, the 
uncertainty on the adopted mean $J-K_{AAO}$ color is at most 0.15 mag.
This means that the uncertainty on the transformed $K_s$-band magnitude is 
0.03 mag. This amount is around half of the mean offset 
(see Fig.~\ref{fig:delta_longmore}), thus supporting the idea to set the 
same $J-K_{AAO}$ for all RRLs
(see Fig.~\ref{fig:delta_longmore}), so it is fine to
set the same $J-K_{AAO}$ for all RRLs. \citet{sollima2004} 
provides the offset of their photometry---the same used
in \citet{sollima06a}---with the 2MASS photometric system: 
$\Delta J$=0.00$\pm$0.10 mag and $\Delta K_s$=--0.04$\pm$0.10 mag.
We adopt these corrections to derive the offset from our mean magnitudes.
The VISTA-system mean magnitudes of \citet{navarrete17} were transformed 
adopting the equations provided by CASU 
(\url{http://casu.ast.cam.ac.uk/surveys-projects/vista/technical/photometric-properties},  
\citealp{fernandezgonzalez2018}).

The systematic offset between the photometry of \citet{longmore1990}
and ours is 0.067$\pm$0.036 mag. We point out that their estimate of the 
distance modulus (13.61 mag) is among the smallest 
in the literature. If we assume
that our photometry is more accurate and correct their distance modulus 
by the quoted offset, we obtain 13.68 mag, which is 
much closer to the bulk of other distance modulus estimates, especially those derived
from RRLs (see \S~\ref{chapt_distance_omega}). 
The work of \citet{longmore1990} was only focused---as
is clear in Fig.~\ref{fig:delta_longmore}---on RRLs far 
from the center of the cluster. Therefore, it is unlikely that the 
offset is due to blended sources in their photometry. A more
plausible explanation is that the standard stars for the 
calibration of the AAO \citep{allencragg1983} have $K$-band 
magnitudes all between 1.5 and 6.0 mag, that are much brighter 
than any RRL in $\omega$ Cen. Moreover, only four of these stars 
were retrieved in the 2MASS catalog by \citet{carpenter2001} to 
derive the AAO-2MASS transformations. Ten more stars from 
\citet{elias1983} were used but these were not considered to be primary standards. 
We conclude that the offset that we found is most likely due to a
combination of an inaccurate calibration---even if it was the 
best possible one---and to a non-precise and, possibly, inaccurate 
transformation between the AAO and 2MASS system.

The comparison with \citet{sollima06a} gives small 
median offsets both in the $J$ ($\sim$0.02 mag) 
and in the $K_s$ band ($\sim$0.01 mag).
However, the dispersion of the magnitude offsets is large ($\sim$0.10 
and $\sim$0.08 mag in $J$ and $K_s$ band, respectively) compared to the 
median (see Fig.~\ref{fig:delta_sollima}). We can safely assume the 
overall offset to be null in both bands. The large sigma is most likely
due to the paucity of phase points that were collected for that 
program, as it was not primarily conceived to obtain RRLs time-series data.

Fig.~\ref{fig:delta_nava} shows the comparison of our mean magnitudes
with those by \citet{navarrete17}. The median offset is $\sim$0.03 mag 
in both $J$ and $K_s$. However, we note that, in the $J$ band, there 
is a trend with distance in the innermost region of the cluster: the closer
the star to the cluster center, the brighter the mean magnitudes of
\citet{navarrete17} compared to ours. This is probably due to better angular
resolution of blended sources in our photometry. This hypothesis is also
supported by the difference in the pixel scales of FourStar 
(0.16 arcsec/pixel) and VIRCAM (0.34 arcsec/pixel). Note that 
the rise in the offset of the $K_s$ magnitudes at 
distances from the center larger than 20 arcmin, is only apparent 
and no firm conclusion can be derived, due
to a paucity of RRLs in these cluster regions.

\section{NIR pulsation properties of RR Lyrae variables}\label{par:rrl_properties}

\subsection{RRL instability strip in NIR and in optical-NIR CMDs}\label{par:rrlcmd}

\begin{figure*}[!htbp]
\centering
\includegraphics[width=13cm]{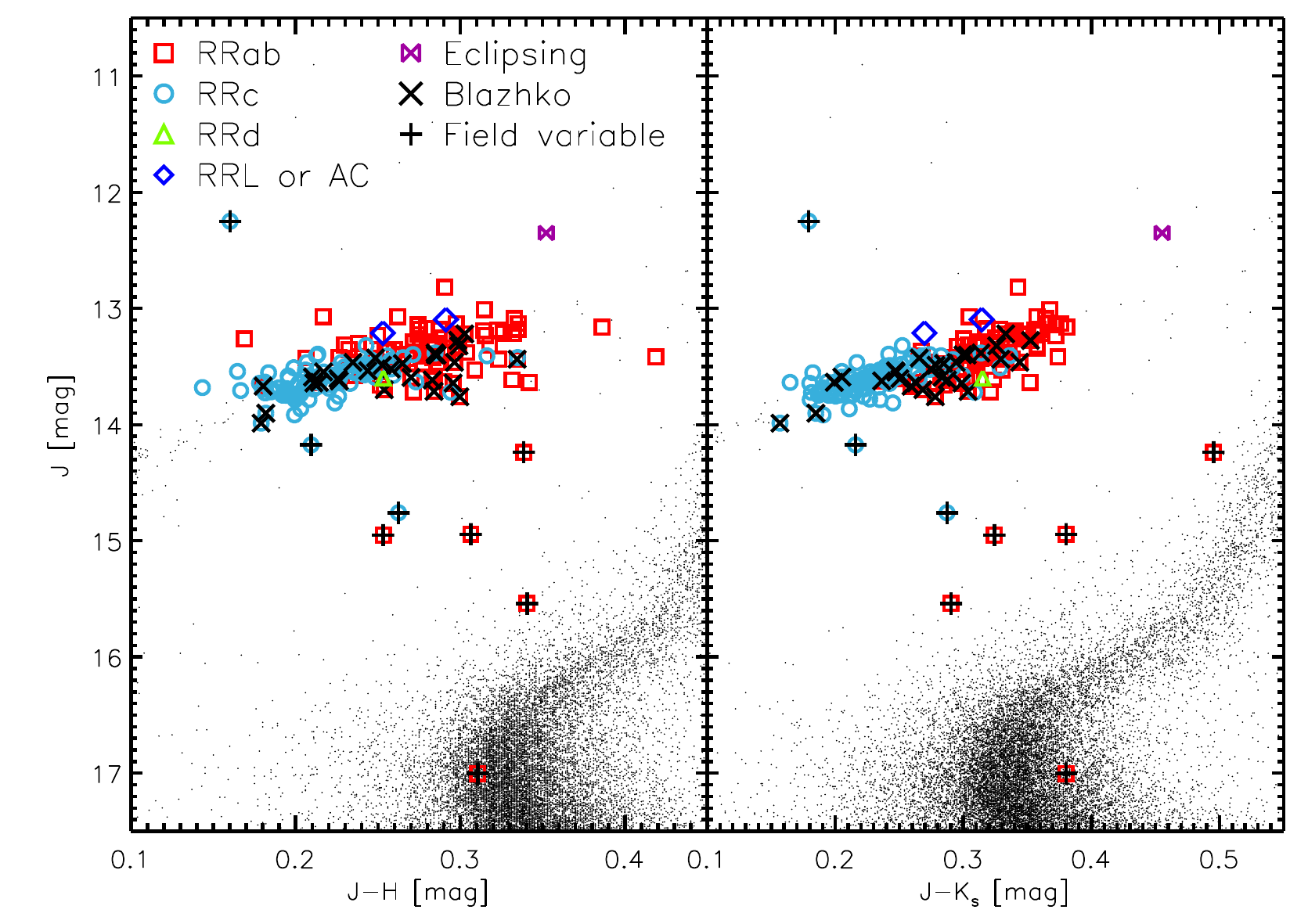}
\caption{Left: NIR ($J$ vs $J-H$) Color-Magnitude Diagram (CMD) of $\omega$ Cen 
(black dots, LCO13 data set). 
Symbols are the same as in Fig.~\ref{fig:delta_spline_poly}.
The candidate RRd variable---V142---is marked with a green triangle, 
the eclipsing binary V179 is marked with a purple bowtie, 
while the black crosses display candidate Blazhko variables. 
Field variables are marked with a black plus. 
Right: same as the left, but for $J$ vs $J-K_s$ CMD}. 
\label{fig:omegacen_cmd_nir}
\end{figure*}

Fig.~\ref{fig:omegacen_cmd_nir} shows the entire set 
of candidate cluster RRLs in $\omega$ Cen in the 
$J$,$J-H$ and in the $J$,$J-K_s$ CMDs. The properties of field 
variables and misclassified variables are discussed in more detail in 
Appendix~\ref{individualnotes}. In this context we note only that candidate field variables are 
located at radial distances larger than 9 arcmin from the cluster center. 

The typical uncertainties of the mean NIR magnitudes 
are at most of the order of a few hundredths of a magnitude. This indicates 
that the range in magnitude covered by the candidate cluster RRLs is 
intrinsic. 

\begin{figure*}[!htbp]
\centering
\includegraphics[width=14cm]{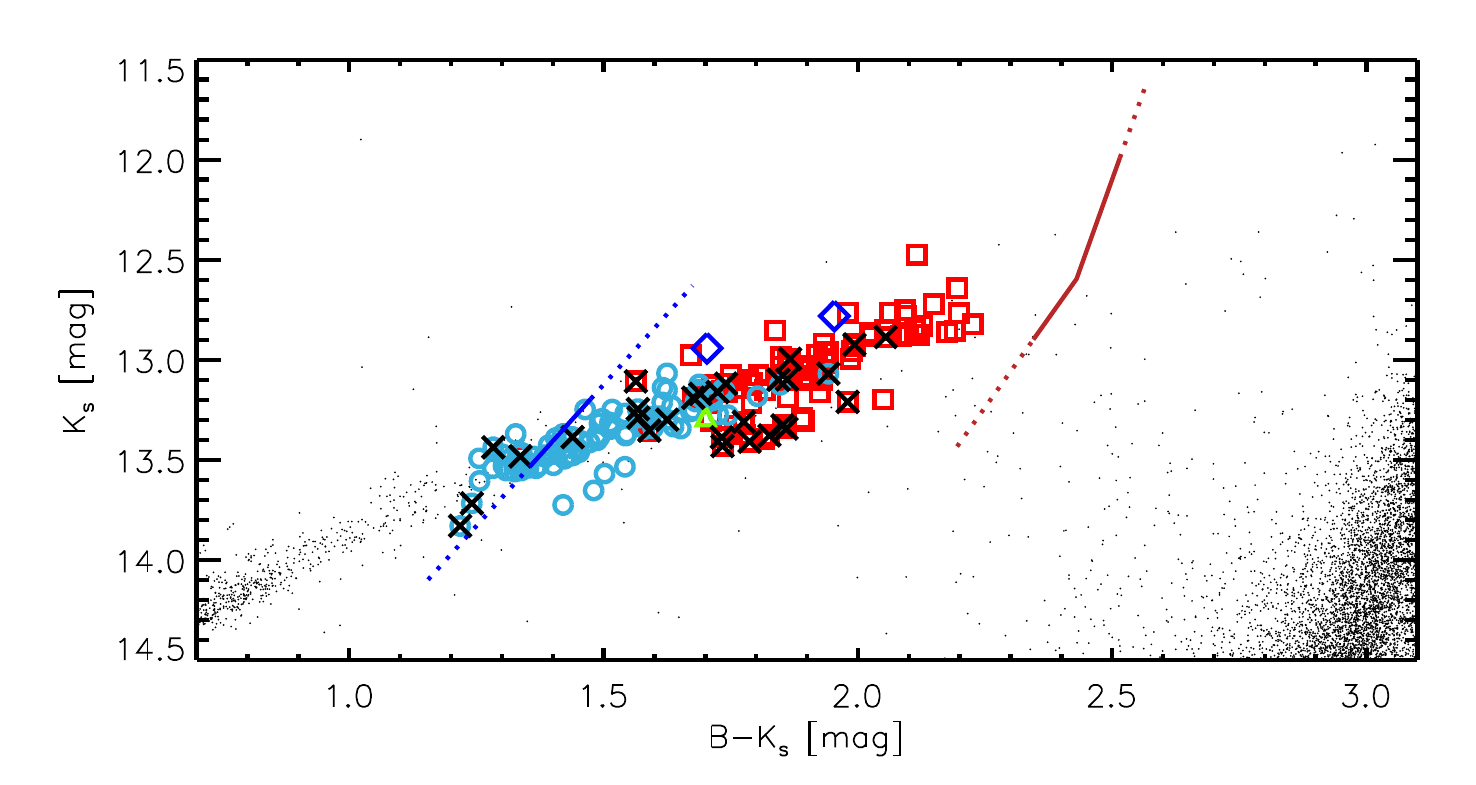}
\caption{Close-up of the RRL instability strip of the 
Optical-NIR ($K_s$ vs $B-K_s$) color-magnitude diagram 
of $\omega$ Cen. The optical photometry for both static and variable stars 
comes from \citet{braga16}, while the NIR is based on the LCO13 data set.
The almost vertical blue and red lines display the predicted first overtone blue edge
and the fundamental red edge according to \citet{marconi15}, at 
Z=0.0006 ([Fe/H] = --1.84; $\alpha$-enhanced mixture).}
\label{fig:omegacen_cmd_kbk_hb}
\end{figure*}

To further improve the analysis of the candidate cluster RRLs 
inside the instability strip, we adopted an optical-NIR CMD 
(Fig.~\ref{fig:omegacen_cmd_kbk_hb}). The color sensitivity increases by 
almost a factor of four (0.2-0.3 vs 1.1 mag) when moving from NIR to optical-NIR 
CMDs. The increased sensitivity in effective temperature provides the opportunity 
to trace the topology of the instability strip. Indeed, the first overtone variables are 
systematically bluer/hotter than fundamental pulsators. They overlap in a region 
called the ``OR'' region \citep{vanalbada73,bono1993,bono97d,caputo1998}, 
i.e., the region in which they can pulsate either 
in the fundamental or in the first overtone mode. The predicted topology is further 
supported by the new mixed-mode candidate identified by \citet{braga16}
even if the actual pulsation mode distribution in the OR region is related 
to the so called ``Hysteresis mechanism'' \citep{vanalbada73,bono95b,bono97d}.

The candidate cluster variables display a narrow distribution 
in magnitude and color. The increased sensitivity of 
the optical-NIR magnitudes allows us to investigate 
in more detail the location of candidate Blazhko RRLs. 
Data plotted in Fig.~\ref{fig:omegacen_cmd_kbk_hb} display that 
they are mainly concentrated in the 
transition between RRc and RRab, as recently suggested by \citet{braga16}.
This finding supports previous results by \citet{jurcsik2011} concerning 
Blazhko RRLs in the Galactic globular M5. Furthermore, it is suggesting that 
the iron content, in the metallicity range covered by RRLs in $\omega$ Cen, 
is playing a marginal role in the transition from RRc to RRab. 
Moreover, there is mounting evidence that they cover 
the entire range of colors typical of RRc variables. 
This empirical evidence is
consistent with recent investigations of Blazhko variables in the Galactic
Bulge \citep{prudil2017}. Interestingly, at $\sim$0.7 days, the NIR/optical 
photometric amplitude ratios display an increase compared to short-period
RRLs (see \S~\ref{par:ampl}).
These findings need to be cautiously treated. The referee found, by using 
optical photometry by \citet{kaluzny04}, that the variable V38 (P=0.779 
days) is indeed a low-amplitude candidate Blazhko RRL. Moreover, she also suggested that 
the detection of Blazhko RRLs in the long period tail is very difficult 
because the amplitude modulation steadily decreases. 
In passing we note that a similar evidence was recently brought forward by \citet{jurcsik2018}
using Galactic bulge RRab variables. These are very interesting findings worth being
investigated in globulars with sizable samples of RRLs, since the age and the chemical
composition of their progenitors is well defined.

The same figure also shows reasonable agreement between the predicted 
first overtone blue edge (blue solid line) plus fundamental red edge (red 
solid line) provided by \citet{marconi15} and observations. However, the 
predicted edges are slightly redder compared to optical-NIR 
observations. The difference is of the order of $\Delta(B-K_s)\sim$0.10-0.15 mag, 
meaning a difference in effective temperature of the order of $\sim$250 K.  
In the comparison between theory and observation we must take account of 
several basic assumptions.  The predicted edges were estimated assuming a 
fixed metal content, Z=0.0006 and an $\alpha$-enhanced mixture, which means [Fe/H] = --1.84 
\citep[similar to the peak in metallicity found by ][]{johnson_e_pilachowski2010}. 
Furthermore, the analytical relations used to estimate the position of the edges 
are based on pulsation models computed with a resolution in effective temperature 
of $\pm$50 K. In a forthcoming investigation we also plan to study the 
impact that stellar atmosphere models have in transforming pulsation predictions 
into the observational plane \citep{marconi15}. On the observational side, 
we are assuming the same cluster reddening for the entire sample of RRLs, 
but there is evidence of reddening variation ($\Delta(B-V)\sim$0.03 mag) 
across the body of the cluster \citep{calamida05}.

It is worth mentioning that candidate RRLs are far from being uniformly 
distributed in color and magnitude across the instability strip. 
\citet{castellani2007}, by using a photometric catalog based both on space 
(\hst) and on ground-based wide-field images, found that the luminosity function 
either in magnitude or in color of HB stars have a clumpy distribution. This 
means that the number of hot HB stars per magnitude interval is not uniform, 
since they display well-defined peaks and gaps. The same outcome applies to the 
instability strip, indeed, the width in color of the instability strip in the 
$K$,$B-K$ CMD is roughly one magnitude ($B-K\sim$1.2--2.2 mag). The number of 
RRLs located in the bluer region ($B-K<$1.45 mag, 40 objects) is higher than 
in the redder region ($B-K>$1.95 mag, 26 objects). Note that this analysis can 
be barely performed in the $V$,$B-V$ CMD, since the width in color of the 
instability strip is only 0.4 mag ($B-V\sim$0.2--0.6 mag).
The current photometric accuracy and the homogenous estimates of both optical 
and NIR mean magnitudes bring forward the occurrence of multiple sequences not 
only among RRab variables, but also among RRc variables. 
This evidence suggests that the metallicity distribution of RRLs in $\omega$ Cen 
is indeed multi-modal. Unfortunately, the accuracy of the current spectroscopic 
and photometric metallicity estimates do not allow us to constrain, on a 
quantitative basis, the magnitude-metallicity variation.


\subsection{Bailey diagram and photometric amplitude ratios}\label{par:ampl}

\begin{figure*}[!htbp]
\centering
\includegraphics[width=8cm]{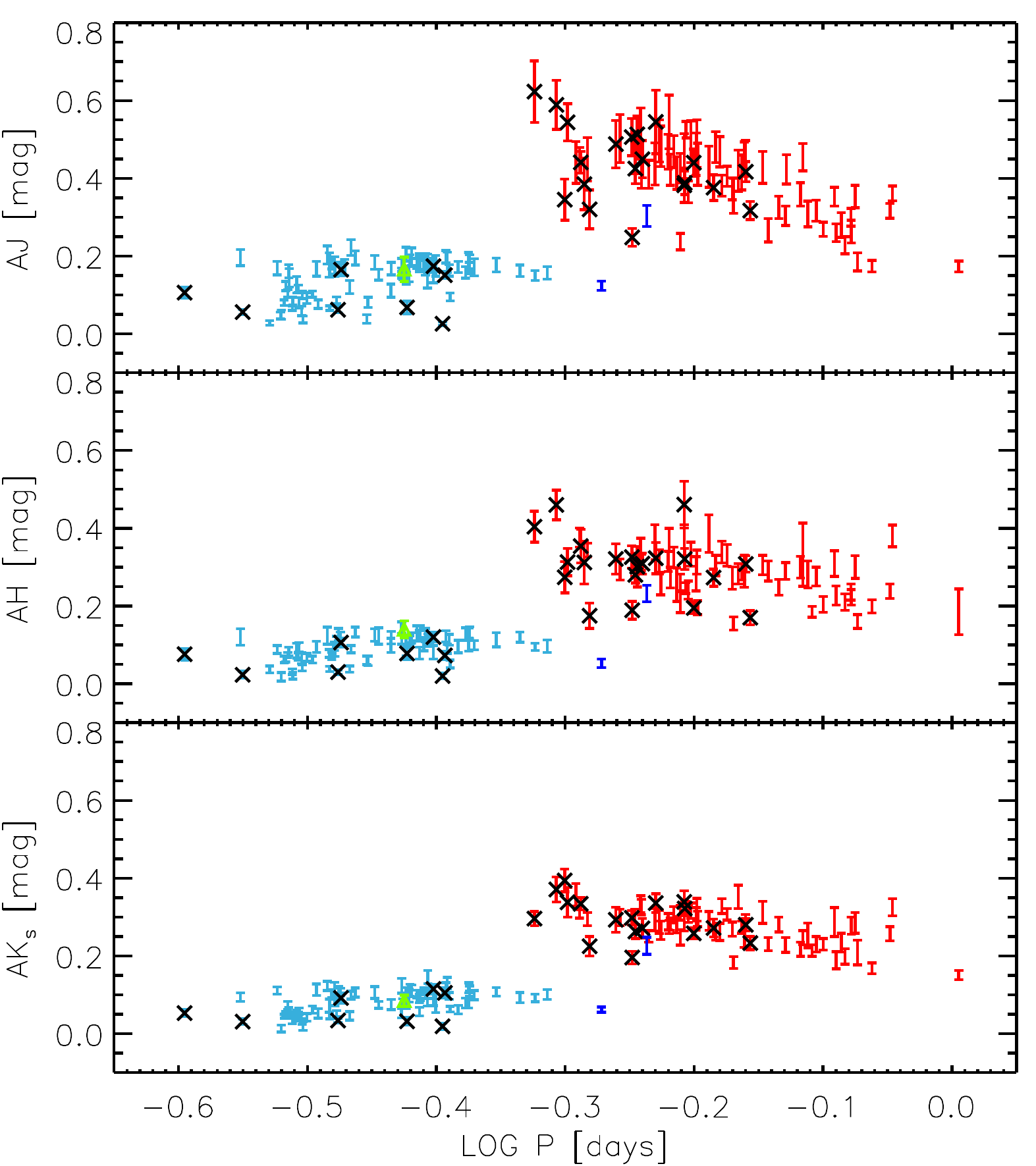}
\caption{Top: $J$-band photometric amplitude versus logarithmic 
period---Bailey Diagram---for $\omega$ Cen RRLs. 
Uncertainties on the amplitudes are shown as vertical error bars.
The symbols and the color coding is the same as in Fig.~\ref{fig:omegacen_cmd_nir}.
Middle: Same as the top, but for $H$-band amplitudes.
Bottom: Same as the top, but for $K_s$-band amplitudes.}
\label{fig:omegacenbailey}
\end{figure*}

\begin{figure*}[!htbp]
\centering
\includegraphics[width=9cm]{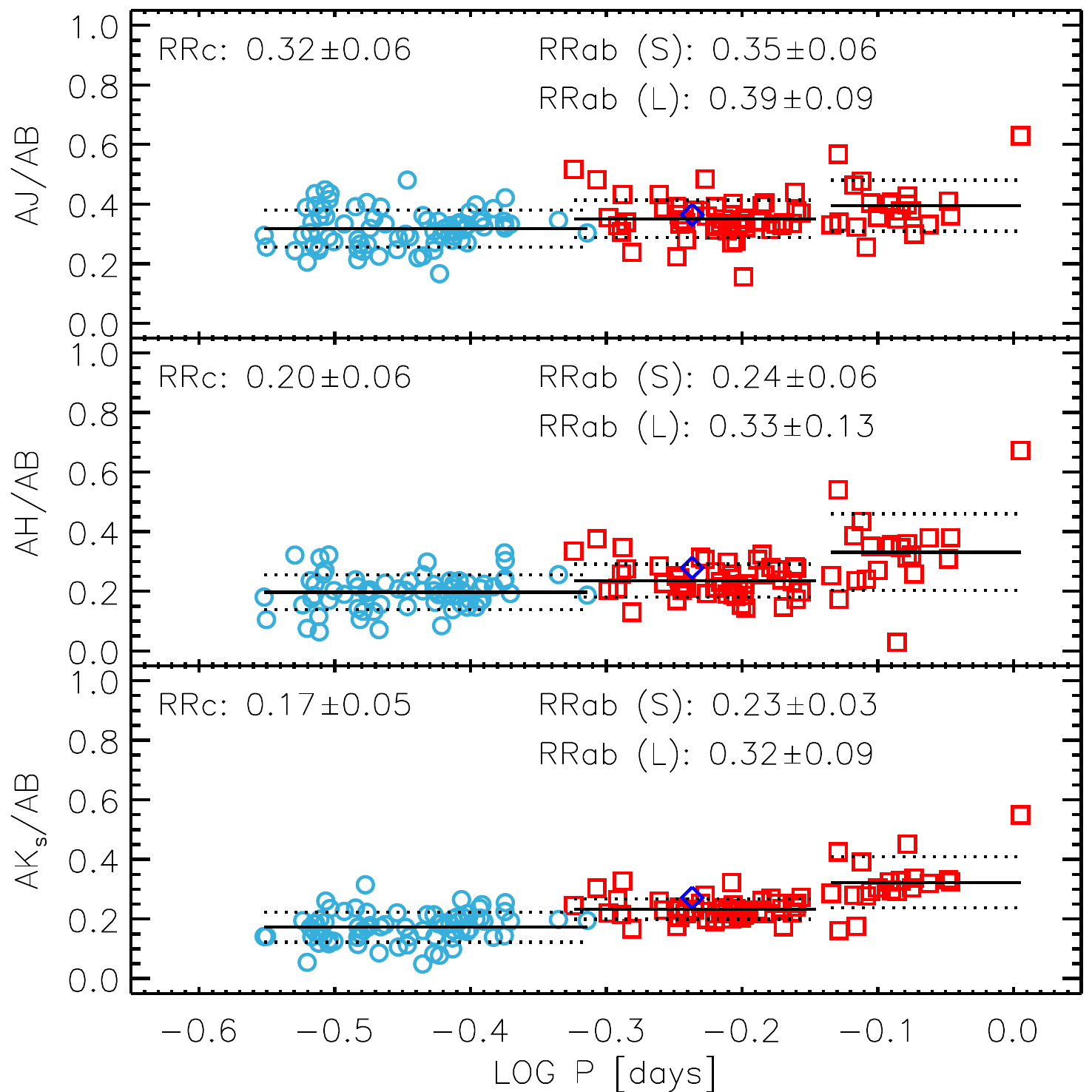}
\caption{Top: NIR/Optical ($AJ$/$AB$) amplitude ratios as a 
function of the logarithmic period. 
The median and the standard deviation of the
RRc sample, of the short-period ($\log P \le -0.15$)
RRab and of the long-period RRab ($\log P > -0.15$)
are plotted as solid and dotted horizontal lines 
(see also labelled values). 
Middle: Same as the top, but for $AH$/$AB$ amplitude ratios. 
Bottom: Same as the top, but for $AK_s$/$AB$ amplitude ratios.}
\label{fig:amplratio_b}
\end{figure*}

\begin{figure*}[!htbp]
\centering
\includegraphics[width=9cm]{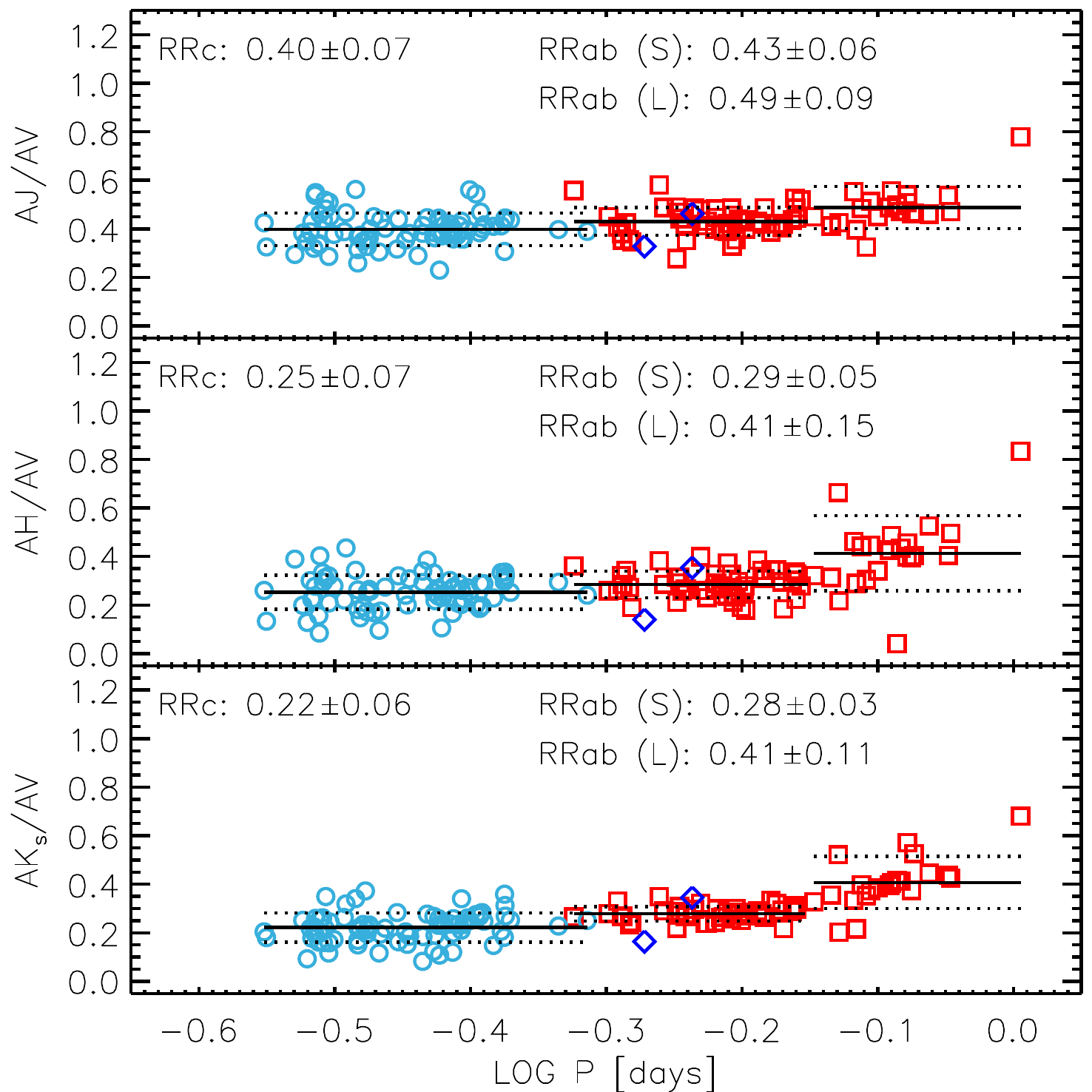}
\caption{Same as Fig.~\ref{fig:amplratio_b}, but the amplitude 
ratios are among NIR bands and the visual bands: $AJ$/$AV$ (top),  
$AH$/$AV$ (middle) and $AK_s$/$AV$ (bottom).}
\label{fig:amplratio_v}
\end{figure*}

\begin{figure*}[!htbp]
\centering
\includegraphics[width=9cm]{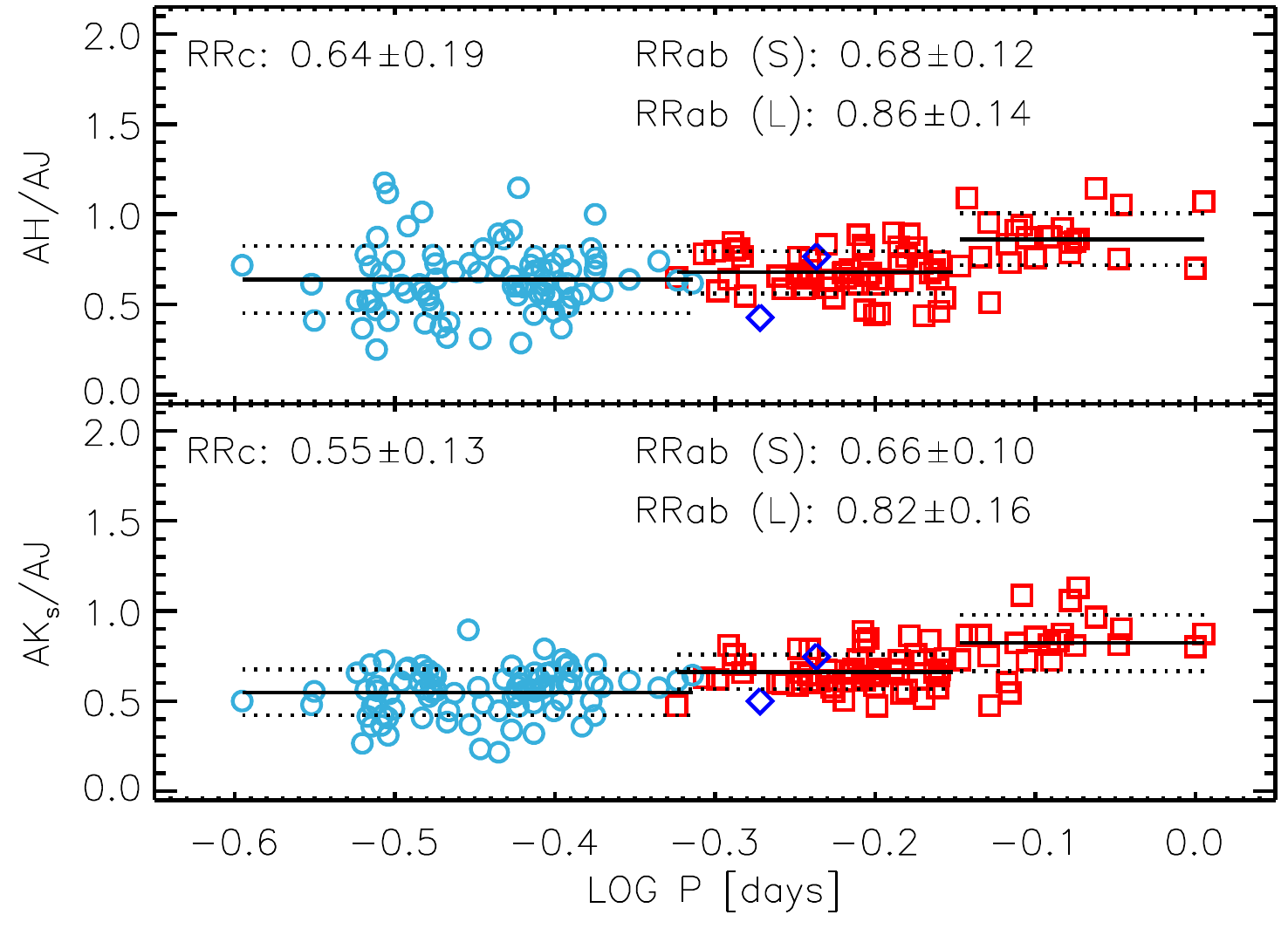}
\caption{Same as Fig.~\ref{fig:amplratio_b}, but for 
the NIR amplitude ratio $AH$/$AJ$ (top) and $AK_s$/$AJ$ (bottom).}
\label{fig:amplratio_jk}
\end{figure*}

The amplitudes of long-period (P$\gtsim$0.7 days) RRab show an almost flat
distribution in the NIR Bailey diagrams (Fig.~\ref{fig:omegacenbailey}). 
This is different from the optical Bailey diagrams, where the amplitudes
steadily decrease toward longer periods \citep{braga16}. 
To better understand the different behavior of optical and NIR amplitudes, 
we have calculated the NIR-over-optical and the $H,K_s$-over-$J$-band
light-amplitude ratios of the RRLs in $\omega$ Cen and we 
have divided the candidate RRLs into three groups: RRc, 
short-period RRab (P$\ltsim$0.7 days) and long-period RRab (P$\gtsim$0.7 days).
The results are shown in 
Figs.~\ref{fig:amplratio_b},~\ref{fig:amplratio_v} and \ref{fig:amplratio_jk}.
 
In Table~\ref{tab:amplratio}, we show the median light-amplitude ratios 
of the RRc, short-period RRab and long-period RRab. 
Note that we use the median instead of the 
mean and that the outliers are ruled out, because these ratios 
will be crucial to derive new light-curve templates and one
must be as accurate as possible to avoid systematic errors.

We found that the median of the amplitude ratios
of RRc is systematically smaller than that of the short-period
RRab, whatever the combination of NIR-to-optical 
photometric bands. Note that this behavior is more and more
evident moving from $J$ (mild increase of amplitude ratios with
period) to $K_s$. As a consequence, the increase 
with period is also seen in the $J$-over-$K_s$ amplitude ratio.
This behavior is different from what was found for
the optical-to-optical amplitude ratios \citep[constant
over all the periods and pulsation modes, ][]{braga16}.

This is the first time that a clear difference, based on 
a large statistical sample, between the 
NIR-over-optical amplitude ratios of RRab and RRc is found. 
An analogous result was obtained by \citet{inno15}, who found 
a dichotomy in the NIR-over-optical amplitude ratios of 
Cepheids in the Magellanic Clouds (MCs), and Milky Way. 
They found that short-period (P$<$20 days) Cepheids 
have amplitude ratios that are smaller than
long-period Cepheids. A detailed study of the amplitude
ratios of RRLs in several GGCs was done by \citet{kunder13}, but they
used only optical data and found no statistically relevant
dichotomy in the amplitude ratios of RRab and RRc.\\

We have also found that RRab variables, for periods longer than 
$\sim$0.7 days, display a well defined increse in the optical/NIR amplitude 
ratios. In a recent investigation, \citet{jurcsik2018}
found evidence of a linear increase in the $AK$ over $AI$ amplitude ratio 
of Galactic bulge RRab variables, when moving from short to long periods.
It is not clear whether the difference in the two behaviours is either 
intrinsic (e.g., due to the metallicity distribution)
or the consequence of a selection bias.

We have also found that, at periods longer than 
$\sim$0.7 days, the ratios with $AH$ and $AK_s$ in
the denominator are no longer 
constant and increase more or less linearly with 
increasing $\log(P)$. This trend is absolutely unique
and has never before been found even for pulsating variables of other types
(SX Phe, Classical and Type II Cepheids).\\

\section{Period-Luminosity relations}\label{chapt_empirical_pl_omega}

\begin{figure*}[!htbp]
\centering
\includegraphics[width=11cm]{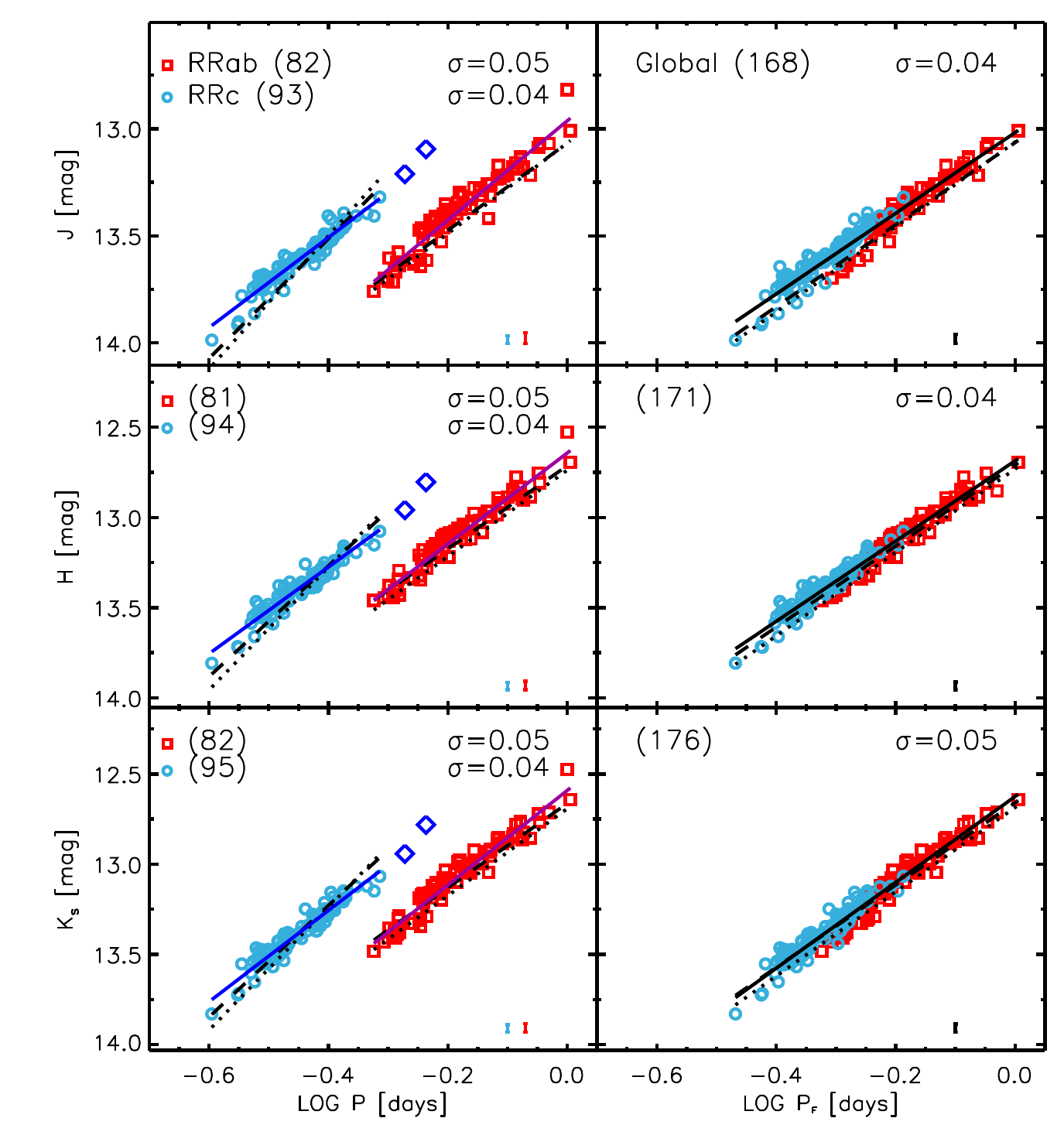}
\caption{Top left: $J$-band Period-Luminosity (PL) relations for 
fudamental (RRab) and first overtone (RRc) RRLs in $\omega$ Cen. 
The black solid lines display the linear fit for 
RRc and RRab. The number of RRab and RRc variables 
that passed the sigma clipping (2.8$\sigma$) is labelled. 
The error bars in the bottom-right corner display the standard 
deviations (see also labelled values). The black
dotted/dashed lines display the theoretical 
PL relations for RRc and RRab variables according 
to \citet{marconi15}, obtained with/without 
including the solar metallicity models in the 
derivation of the theoretical PLs. The predicted relations 
were plotted assuming a true distance modulus of 13.71 mag 
\citep{braga16} and a mean reddening of 0.11 mag \citep{thompson2001,lub2002}.  
Top-right: same as the left, but for the Global sample, i.e. the periods 
of RRc variables were fundamentalized: $\log P_F$=$\log P_{FO}$+$0.127$.
Middle: same as the top, but for the $H$-band PL relation.
Bottom: same as the top, but for the $K_s$-band PL relation.}
\label{fig:pl_jhk}
\end{figure*}

The current NIR data set provided an opportunity to 
investigate in detail the NIR PL relations.

{\em Spread and slope of the PL relations}--Data plotted in 
Fig.~\ref{fig:pl_jhk} show that the standard deviations of RRab (red squares, 
left panel) RRc (light blue circles, left panel) and Global 
(right panels)\footnote{This sample includes both RRab and RRc variables. 
The periods of the latter were fundamentalized, i.e. 
$\log P_F$=$\log P_{FO}$+0.127.} variables remain almost constant
(0.04-0.05 mag) when moving from the $J$ to the $K_s$ band. 
This is a significant advantage when compared 
with the standard deviations in the I-band 
(Global, 0.06-0.08 mag) of the same variables \citep{braga16}. The same outcome 
applies to the slopes of the PL relations, indeed they increase for the RRc 
from --2.105 ($J$) to --2.531 ($K_s$), for the RRab from --2.318 ($J$) to --2.621 
($K_s$) and for the Global sample from --1.884 ($J$) to --2.380 ($K_s$). The
improvement is even more relevant when compared with the slopes for 
the $I$-band (RRc, --1.624; RRab, --1.955; Global, --1.335).


{\em Metallicity dependence}--We performed a linear fit over the entire 
data set and the resulting coefficients are listed in Table~\ref{tab:pl}. 
The errors on the mean magnitudes are on average of the order of a few 
hundredths of a magnitude (see Table~\ref{tab:pl}). This indicates that 
the dispersion along the PL relations and the errors on the slopes are 
dominated by an intrinsic feature. In particular, several variables are 
intrinsically either brighter or fainter than the linear fit. This suggests 
that they are either more metal-poor or more metal-rich than the bulk of 
RRL variables. This further supports recent findings concerning the metallicity 
dependence of the RRL PL relations based on theoretical \citep{marconi15} and on 
semi-empirical \citep{neeley17} arguments. 

The peak of the metallicity distribution in $\omega$ Cen is 
[Fe/H]$\approx$--1.8 \citep{johnson_e_pilachowski2010}. To provide more 
quantitative constraints, we performed a detailed comparison with predicted 
NIR PL relations recently provided by \citet{marconi15}. In this investigation 
the authors provided NIR PL relations that either neglect (metal-independent) 
or take account of the metallicity dependence. The key advantage of the quoted 
predictions is that they cover a broad range in stellar masses, luminosity 
levels and chemical compositions.
In particular, they adopted an $\alpha$-enhanced chemical composition, a 
constant helium-to-metal enrichment ratio and seven different metal abundances 
ranging from Z=0.0001 to Z=0.0198. This means that the predictions cover more 
than two dex in iron abundance. The dotted lines plotted in
Fig.~\ref{fig:pl_jhk} display the metal-independent PL relation, and the 
agreement for both RRab, RRc and Global variables is quite good, if we take 
account for the significantly different metallicity distributions covered by 
theory and observations. Indeed, once predicted metal-independent PL 
relations are restricted to models more metal-poor than solar 
(Z$\leq$0.008), the predicted relations display zero-points and slopes that 
agree better with the observed ones (see dashed lines on the same figure).

The above circumstantial evidence indicates that 
NIR PL relations depend on metallicity. 
To further investigate this effect we computed new 
empirical Period-Luminosity-Metallicity 
(PLZ) relations using three different sets of iron abundances: 
1) spectroscopic estimates for 74 RRLs provided by \citet{sollima06a}; 
2) photometric estimates based on $Ca,b,y$ photometry provided by 
\citep{rey2000} for 131 RRLs;   
3) photometric estimates obtained by \citet{braga16} by inverting the $I$-band 
PLZ relation for 160 RRLs.   

\begin{figure*}[!htbp]
\centering
\includegraphics[width=16cm]{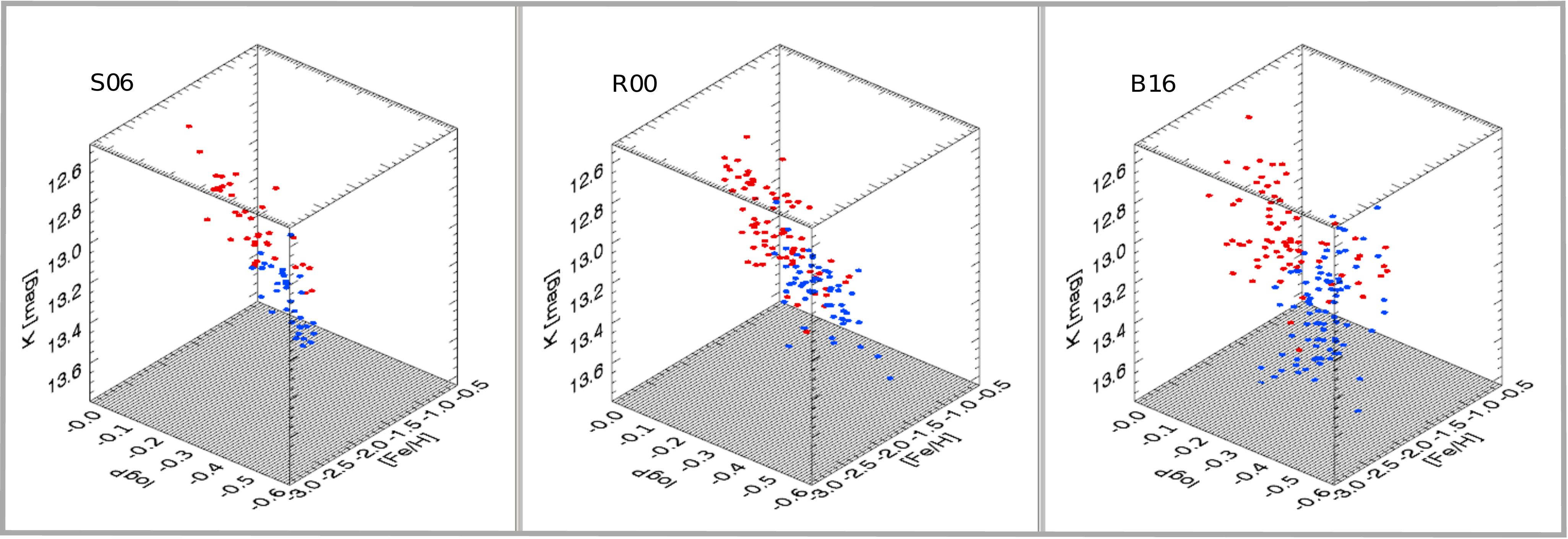}
\caption{Left: $K_s$-band Period-Luminosity-Metallicity relation for RRLs 
in $\omega$ Cen. The photometric metallicity ([Fe/H]) estimates come 
from \citet{sollima06a}. The symbols and the color coding are the same 
as in Fig.~\ref{fig:omegacen_cmd_nir}.
Middle: Same as the left, but spectroscopic metal abundances according to \citet{rey2000}.  
Right: Same as the left, but photometric metallicity estimates according to 
\citet[][Table 10, fifth column]{braga16}.}
\label{fig:pl_3d}
\end{figure*}

The coefficients of the surface fits and their 
uncertainties are listed in Table~\ref{tab:plz}, 
while Fig.~\ref{fig:pl_3d} shows the 
NIR PLZ relations in 3D plots. There is evidence that the spread in magnitude is, 
at fixed period, dominated by the spread in metal abundances. The metallicity 
estimates are affected, on average, by uncertainties of the order of 0.2 dex.    
The difference among the three different sets of metal abundances is caused 
by the sample size; indeed the metallicity estimates by \citet{braga16} are more 
than a factor of two larger than the spectroscopic sample by \citet{sollima06a}. 
The sample size can play a crucial role in constraining the spread in metallicity. 
Short-period RRc variables are marginally present in the \citet{rey2000} sample, but they 
appear in \citet{sollima06a} and in \citet{braga16}. The current data suggest 
that they fix the metal-poor tail of the sample.  

\begin{figure*}[!htbp]
\centering
\includegraphics[width=9cm]{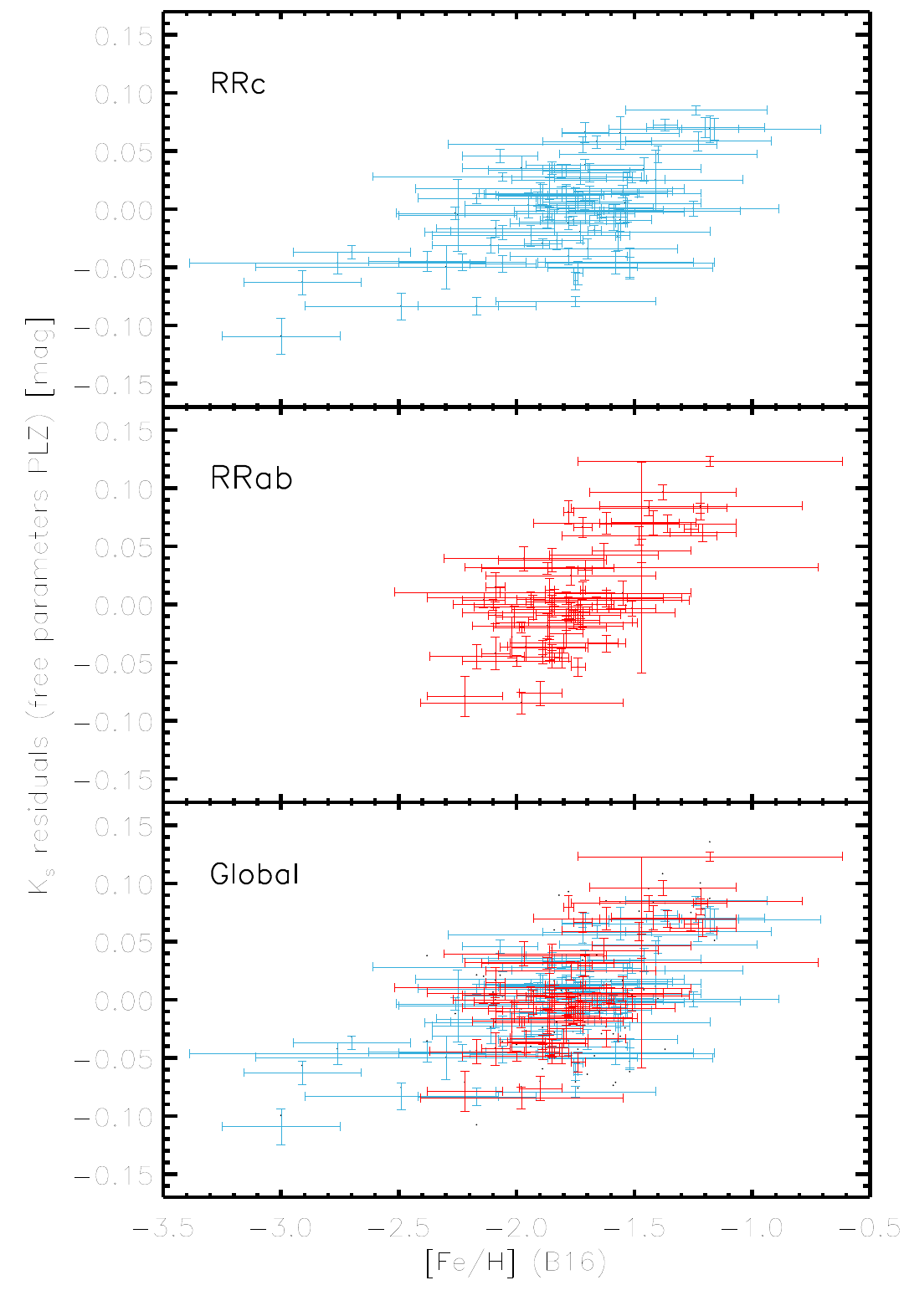}
\caption{Top: Residuals of the $K_s$-band mean magnitudes of
RRc stars minus the empirical PLZ---derived by leaving free 
all the parameters and using metallicities by
\citet{braga16}---at fixed [Fe/H]=--1.8 dex, versus iron abundance.
Middle: Same as the top, but for RRab stars.  
Bottom: Same as top, but for the Global sample.}
\label{fig:pl_residuals}
\end{figure*}

To show even more clearly 
the dependence on metallicity of the PL relations, we also display, 
in Fig.~\ref{fig:pl_residuals}, the residuals of the $K_s$-band mean magnitudes 
with respect to the PLZ at fixed iron abundance ([Fe/H]--1.8 dex). 
This metal abundance was selected because it is the peak of the RRL metallicity 
distribution \citep{braga16}. The residuals display a clear trend in spite of the 
large dispersion. This means that the lack of a metallicity term in the PL relations 
leads to magnitudes that are systematically brighter/fainter for 
metal-poor/metal-rich RRLs.

\begin{figure*}[!htbp]
\centering
\includegraphics[width=11cm]{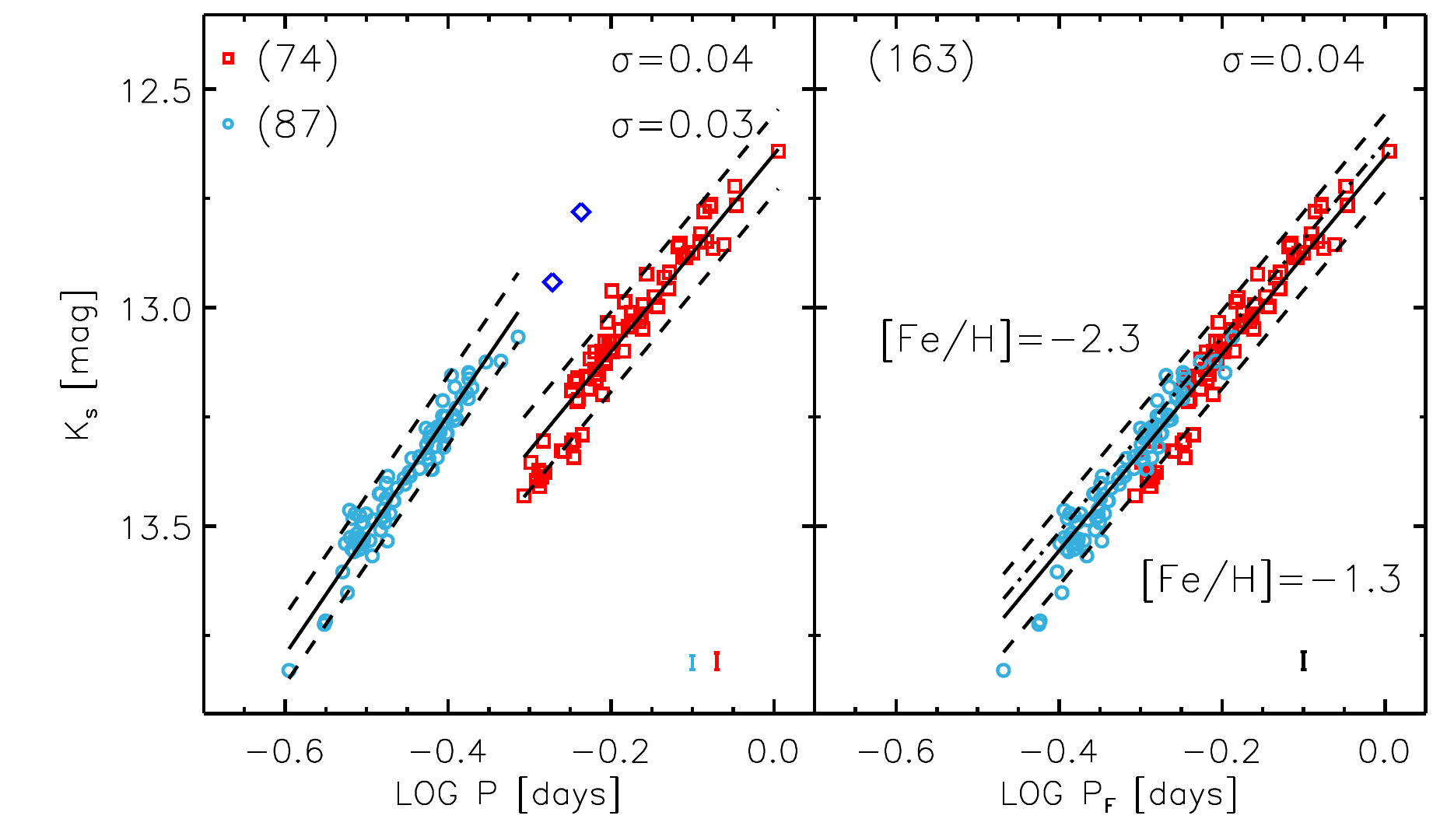}
\caption{Left: $K_s$-band Period-Luminosity relation for RRab and RRc in 
$\omega$ Cen. The black dashed lines display the predicted PLZ relations 
by \citet{marconi15} at two fixed metal abundances: 
[Fe/H] = --1.3 (fainter) and [Fe/H] = --2.3 (brighter). 
Theory was plotted assuming a true distance modulus of 13.71 mag
and a mean cluster reddening of 0.11 mag \citep{thompson2001,lub2002}.
The black solid lines display the surface fit to the data---the coefficients 
are listed in Table~\ref{tab:plz}---by assuming a fixed metal 
abundance ([Fe/H]=--1.8) and the slopes predicted from theory.
The number of variables adopted in the fit and the 
standard deviations from the PLZ relations are 
also labelled. The error bars in the bottom-right corner display 
the standard deviations (see also labelled values).
Right: Same as the left, but for the Global (RRab+RRc) sample. 
The black dotted-dashed line shows the PLZ relation derived by \citep{navarrete17}
assuming a metal content of [Fe/H]=--1.8.}
\label{fig:pl_fixedslope}
\end{figure*}

To further quantify the spread in metal abundance 
of $\omega$ Cen RRLs, Fig.~\ref{fig:pl_fixedslope} shows
the $K_s$-band PL relation together with predicted PLZ relations at fixed metal abundance
(dashed lines). The comparison between theory and observations 
indicates that the RRLs span more than one dex in metal abundance, 
indeed they range from [Fe/H]$\approx$--2.3 to $\approx$--1.3. To further validate    
the quoted trend we decided to perform a linear fit to the mean NIR magnitudes 
using the predicted slope for the period dependence and at fixed metal content 
([Fe/H]=--1.8).

The black solid lines plotted in the left panel of this figure show 
very good agreement between theory and observations for both RRc and RRab variables
over the entire period range. 
The outcome is the same for the fundamentalized 
PLZ relation (black solid line in the right panel) 
plotted in the right panel of the same figure. This panel also shows the comparison 
with the empirical PLZ relation recently derived by \citet{navarrete17}. The 
agreement is also reasonable, but it covers the upper envelope of the observed 
distribution. This difference is easily explained by the difference between 
our mean magnitudes and those by \citet{navarrete17}, already 
addressed in \S~\ref{par:magcompare}.
The referee suggested to double check the impact of the individual sigma 
clipping applied to estimate the PLZ relations. We computed a new set of 
PLZ relations using the same sigma clipping that we adopted for the metal-independent 
PL relations. We found that the coefficients of the PLZ relations are, within 
the errors, minimally affected by the different sigma clipping that we applied.

We also note, in Table~\ref{tab:plz}, that the best agreement for the 
metallicity coefficients of the PLZ relations is found for the RRab and the 
Global sample using the iron abundances from \citet{sollima06a}. The metallicity 
coefficients of RRc PLZ relations are almost null when we adopt either the 
\citet{rey2000} or the \citet{sollima06a} iron abundances. On the
other hand, the agreement with the predicted PLZs \citep{marconi15} 
is good when using abundances provided by \citet{braga16} and 
the dispersion among RRc, RRab and Global samples is also smaller. Finally, 
the dispersion of the data around the PLZ is only marginally smaller 
($<$0.01 mag) than the dispersion around the PL relations 
(see Figs.~\ref{fig:pl_jhk} and \ref{fig:pl_fixedslope}). This further supports 
the uncertainties affecting the current spectroscopic estimates.

Figures \ref{fig:pl_jhk} and \ref{fig:pl_fixedslope} also show the 
position of the two variables (V68 and V84) whose 
classification is unclear and that are plotted as blue diamonds.
The classification as cluster ACs can be discarded on the basis of empirical evidence. 
Detailed investigations in nearby dwarf galaxies indicate that ACs are typically 
one magnitude brighter than the bulk of RRLs (Carina, \citealt{coppola15};
Sculptor, \citealt{martinezvazquez16b}; MCs, \citealt{soszynski2015a}). During the last few 
years there is evidence in nearby dwarfs of a few variables of uncertain 
classification, with optical magnitudes between the RRLs and the ACs. However, 
these variables are, on average, at least half a magnitude brighter in $V$ than 
the upper envelope of RRLs. V68 is indeed likely to be a cluster RRc from the 
PL relations, while V84 is more likely to be either a background FO AC or 
a foreground RRab. The matter is discussed in more detail in 
Appendix~\ref{individualnotes}.

The above findings are solid empirical evidence of the metallicity dependence 
of NIR PL relations and calls for more accurate and homogeneous spectroscopic 
abundances for $\omega$ Cen RRLs.

\section{$\omega$ Cen distance determinations}\label{chapt_distance_omega}

There is mounting theoretical and empirical evidence that NIR and MIR 
PL relations of RRLs are solid distance indicators 
\citep{longmore1990,bono03c,catelan04,dallora04,madore13,braga15,neeley17}. 
Moreover, RRLs have the advantage of being old (t$\ge$10 Gyr) stellar tracers: 
this implies the opportunity---through the calibration of
the luminosity of the Tip of the Red Giant Branch (TRGB)---to achieve a homogeneous 
extragalactic distance scale that is independent of classical Cepheids 
and applies to both early and late type galaxies \citep{beaton2016}.
In the calibration of the distance scale based on old stellar tracers 
$\omega$ Cen will play a crucial role. $\omega$ Cen is a unique Galactic 
globular, since it hosts a sizeable sample of RRLs and it is massive 
enough to provide a solid estimate of the TRGB luminosity (the TRGB can also 
be easily measured in 47 Tuc, but this cluster hosts only one RRL). 
The current estimates of the TRGB luminosity of $\omega$ Cen 
in the NIR bands have been provided by \citep[][$J^TRGB$ = 8.59$\pm$0.06, 
$H^TRGB$ = 7.81$\pm$0.08 and $K^TRGB$ = 7.70$\pm$0.06]{bellazzini04}, adopting
the geometrical distance based on eclipsing binaries provided by \citet{thompson2001}.
The TRGB was also used as a standard candle to derive a true distance 
modulus to $\omega$ Cen of $\mu$=13.65$\pm$0.05 mag \citep{bono2008b}.
The added value of $\omega$ Cen as distance calibrator is that Gaia will
provide an accurate geometrical distance, since it is among the 20 
closest (d $\lesssim$ 5--6 kpc) globulars. \\

The NIR time series we collected allowed us to provide accurate 
and precise mean magnitudes, and in turn, accurate distances to 
$\omega$ Cen. To constrain possible uncertainties in the adopted 
zero-point of the NIR PL relations we decided to follow two different 
approaches. 

1) {\em Theoretical calibration} --  
we used the predicted PLZ relations using the pulsation models by \citep{marconi15}, but 
adopting only metal abundances ranging from Z=0.0001 to Z=0.008. 

2) {\em Empirical calibration} -- we adopted the slope and the metallicity 
coefficient of the current empirical PLZ relations (see Table~\ref{tab:plz}) 
and the zero-points based on the five Galactic RRLs (4 RRab, 1 RRc) for 
which \citet{benedict11} provided trigonometric parallaxes using the 
the Fine Guide Sensor available at \hst. 
The ranges in metal abundance (--1.8$<$[Fe/H]$<$--1.4)
and in pulsation period (0.31$<$P$<$0.66 days) covered 
by these five calibrators are modest \citep{benedict11,coppola15}. 
Moreover, it is becoming more evident that both the trigonometric 
parallaxes and the extinctions of these calibrators appear to be affected 
by systematics. The reader interested in a more detailed discussion 
concerning the \hst calibrators is referred to \cite{neeley17}. 

The true distance moduli to $\omega$ Cen based on both theoretical and empirical calibrations 
for RRab, RRc and Global samples are listed in Table~\ref{tab:dmod}. 
Uncertainties affecting the cluster reddening and/or the possible occurrence of 
differential reddening \citep{calamida05} are strongly mitigated in the NIR regime. 
This is the reason why we adopted a mean cluster reddening 
of E($B-V$)=0.11 mag \citep[][and references therein]{calamida05}.
Note that distance moduli are based on the metal abundances provided either 
by \citet{sollima06a} or by \citet{braga16}. We neglected the metallicities 
provided by \citet{rey2000} because they provide vanishing metallicity coefficients 
of the PLZ relations, but this trend appears to be at odds with both theory 
\citep{bono03a,marconi15} and observations \citep{martinezvazquez15,braga16,neeley17}.

The distance moduli listed in the aforementioned table bring forward several interesting 
findings concerning the theoretical calibration. 

{\it i}) The agreement between the distance moduli based on 
three different groups (RRab, RRc, 
Global: RRab$+$RRc) is better than 0.4\% (0.5$\sigma$) adopting both the 
the PL$J$, PL$H$ and the PL$K_s$ relations and any of the 
iron abundances by \citet{sollima06a} and \citet{braga16}.
This supports the accuracy of the time series photometry 
and its absolute calibration.

{\it ii}) Cluster distances are independent of the adopted metallicity distribution, 
and indeed the two overall means agree within $\sim$0.2\% (0.3$\sigma$, 
see values listed in Table~\ref{tab:dmod}).

{\it iii}) The dispersion of the metallicity distribution 
based on photometric indices is roughly a 
factor of three larger than the spectroscopic one. However, the standard 
deviation of the cluster distances based on the former sample is only 10-20\% 
larger than the latter one. Thus suggesting that current cluster distance 
distributions are driven by uncertainties in individual iron abundances.

The cluster distances based on the empirical calibration bring forward the 
following findings. 
 
a) Cluster distances show variations among the different samples (RRab, RRc, Global), 
the three adopted bands and the adopted metallicity distributions. Moreover, the error 
in the mean cluster distance is a factor of two larger than the error based on 
the theoretical calibration. We also found that empirical RRc cluster 
distances agree quite well with similar distances based on the theoretical 
calibration. However, empirical RRab cluster distances are systematically 
larger than empirical RRc ones. The difference is caused by the fact that the 
$JHK_s$ mean magnitudes of RZ Cep---the calibrating RRc star---agree within 0.1 mag 
with the predicted NIR PLZ relations from \citet{marconi15}. On the other hand, 
three out of the four calibrating RRab (all except XZ Cyg) appear to be 
overluminous by 0.1-0.2 mag when compared with predicted magnitudes.
This is the reason why the RRab sample provides cluster distances that are 
more than 0.1 mag larger (2.5$\sigma$).
The Global sample averages the two effects and provides empirical distances 
that are slightly larger than those based on the theoretical calibration 
(1.8$\sigma$). 

b) The standard deviation of the overall means (bold text in Table~\ref{tab:dmod}) 
on cluster distances based on the empirical 
calibration is from 15\% (photometric metallicities) to 20\% (spectroscopic 
metallicities) larger than those based on the theoretical calibrations. 

The above findings further support the results obtained by \citet{neeley17} 
and suggest that the current RRL calibrators are still affected by systematics 
at the 5-10\% level. These are the reasons why we preferentially trust cluster 
distances based on the theoretical calibration, that should be preferred.
%

\begin{figure*}[!htbp]
\centering
\includegraphics[width=11cm]{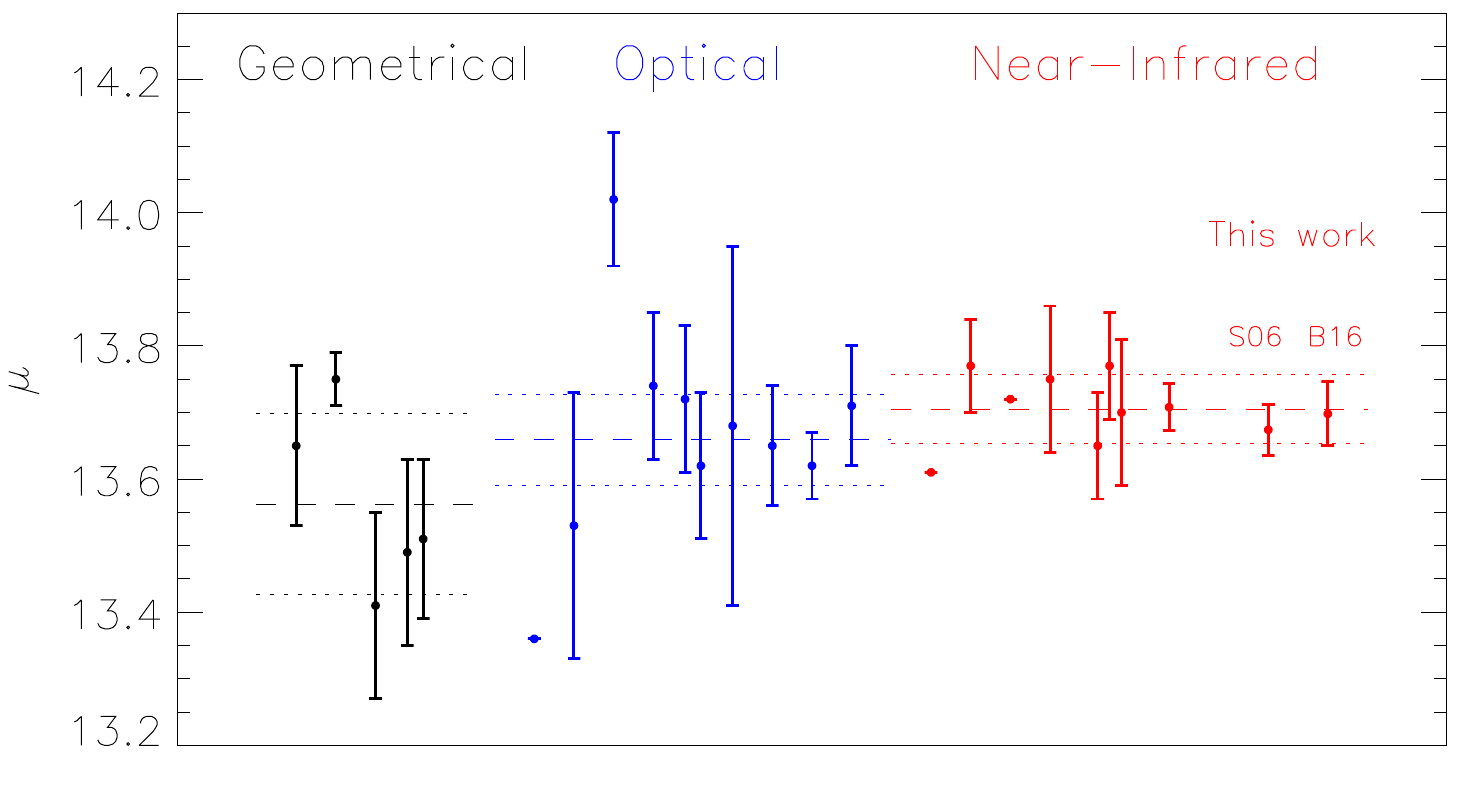}
\caption{True cluster distance moduli to $\omega$ Cen. Black points display 
distance moduli based on geometrical diagnostics, while blue and red 
points display distance moduli based either on optical or on NIR distance
diagnostics. ``S06'' and ``B16'' mark the current overall mean of the 
distance moduli displayed in Table~\ref{tab:dmod} derived from
the theoretical calibration and based on spectroscopic and on 
photometric metallicity distributions, respectively. The horizontal 
dashed lines display the mean true distance modulus based on the three 
different approaches. The dotted lines display the dispersions 
(see Table~\ref{tbl:omegadistance}). Note that distances from 
\citet{cannon1974} and \citet{mcnamara00} were not used to
derive the mean of the optical sample (see text).}
\label{fig:dmod_lit}
\end{figure*} 

A glance at the true distance moduli to $\omega$ Cen available in the literature, 
listed in Table~\ref{tbl:omegadistance} and plotted in Fig.~\ref{fig:dmod_lit}, 
shows that the true distance moduli agree within 1$\sigma$. However, there is 
mounting evidence that cluster distances based on geometrical methods are 
systematically smaller than those based either on optical or on NIR distance 
diagnostics \citep{bono2008b}. The difference with the distance based on 
the cluster kinematics \citep{vandeven2006} is roughly at the 2$\sigma$ 
level. The reasons for this discrepancy are not clear yet.  

The marginal difference between cluster distances based on optical and 
NIR distance diagnostics might be due to different assumptions concerning 
cluster reddening and/or zero-point calibrations. Two cluster distances 
display discrepancies larger than 1$\sigma$: 
a cluster distance based on an old estimate of the HB luminosity level  
\citep[$\mu$=13.36$\pm$0.10 mag,][]{cannon1974} and 
a cluster distance based on the PL relation of $\delta$~Sct stars 
\citep[$\mu$=14.02$\pm$0.10 mag,][]{mcnamara00}. However, this determination 
is strongly affected by the adopted PL calibration \citep{mcnamara1997}. 
More recent and accurate calibrations \citep{mcnamara2004,mcnamara07} 
would provide smaller cluster distances.

Finally, it is worth mentioning that remarkable agreement is found
with literature estimates based on the TRGB \citep[13.65$\pm$0.05 mag,][]{bono2008b} 
and on RRLs (13.77$\pm$0.07 mag by \citealt{delprincipe06}; 
13.71$\pm$0.08$\pm$0.01 mag by \citealt{braga16} and
13.708$\pm$0.035 mag by \citealt{navarrete17}). 

\section{Summary and final remarks}\label{chapt_final}

In this work we have provided 
accurate and homogeneous mean optical and NIR magnitudes for a large, unbiased
sample of RRLs in $\omega$ Cen. The main results obtained from the analysis 
of the data that we have collected are the following.

{\em Completeness of the RR Lyrae sample.} We provide a complete 
characterization of RRL variables in $\omega$ Cen with either  
unknown or uncertain pulsation mode. See V68, V84, V171, V178, 
V179 and V182 in Appendix~\ref{individualnotes} for more details.

{\em Pulsation period.} The time interval covered by our optical and NIR 
time series data allowed the detection of a period variation for 
$\sim$30 RRL variables. The magnitude of these variations typically agrees 
with specific studies on period change rates \citep{jurcsik2001}.

{\em Amplitude ratios.} The ratios of NIR-over-optical light amplitudes 
for RRc variables are marginally smaller than for RRab variables. We also found 
that these ratios are even larger for RRab stars with periods longer than 
0.7 days. Interestingly enough, in $\omega$ Cen, at periods longer than 
0.7 days, there is a paucity of Blazhko RRLs and their amplitude 
modulation become smaller \citep{braga16}. 
It is worth mentioning that the quoted differences 
are within 1$\sigma$.

{\em PLZ relations.} The spread in magnitude is consistent
with a spread in iron abundance of at least one dex. Moreover, 
the dispersion is smaller using PLZ relations than when using PL relations. Both 
are empirical indications that metallicity does affect the 
zero-point of the PL relation of RRLs.

{\em Distance determinations.} We derived the distance modulus to
$\omega$ Cen from the PLZ($JHK_s$) relations adopting both an empirical 
and a theoretical calibration. Moreover, we also adopted two different 
sets of iron distributions \citep{sollima06a,braga16}. True distance moduli 
based on the theoretical calibration give 13.674$\pm$0.008$\pm$0.038 
mag \citep[spectroscopic iron abundances from][]{sollima06a} and  
13.698$\pm$0.004$\pm$0.048 mag \citep[photometric iron abundances from][]{braga16}. 
These estimates agree quite well (1$\sigma$) with recent cluster distance 
determinations based either on optical or on NIR distance diagnostics. 
Cluster distances based on geometrical methods, taken at face value, appear 
systematically smaller than those based on other distance diagnostics.  

True cluster distance moduli based on the empirical calibration have 
errors in the mean that are, on average, a factor of two larger than 
those based on the theoretical calibration. The current findings 
support previous results obtained by \citet{neeley17} that were based 
on field and cluster RRLs. There is mounting evidence that the five 
adopted HST calibrators are affected by systematics at the 5-10\% level.   

This is a long term project aimed at investigating variable stars and 
stellar populations in $\omega$ Cen. We have already performed an optical 
and a NIR analysis of RRL properties. In a subsequent paper we plan to 
investigate the metallicity distribution---by providing accurate 
spectroscopic abundances for a significant fraction of 
RRLs in $\omega$ Cen using high-resolution optical 
spectra---the reddening distribution and the 
distance distribution using both optical and NIR mean magnitudes. 
These are stepping stones for deriving new NIR ($JHK_s$) light curve 
templates.

\acknowledgements 
We thank the referee J.~Jurcsik for her positive and insightful suggestions that 
improved the content and the readability of the paper. It is a pleasure to thank 
J.~L.~Prieto for his help with time series optical data from ASAS-SN. 
This publication makes use of data products from the Two Micron All Sky Survey, 
which is a joint project of the University of Massachusetts and the Infrared Processing 
and Analysis Center/California Institute of Technology, funded by the National 
Aeronautics and Space Administration and the National Science Foundation.
This publication makes use of data products from the All-Sky Automated Survey for Supernovae
that is supported by the Gordon and Betty Moore Foundation through grant 
GBMF5490 to the Ohio State University and NSF grant AST-1515927.
This research has made use of the USNO Image and Catalogue Archive
operated by the United States Naval Observatory, Flagstaff Station
(\url{http://www.nofs.navy.mil/data/fchpix/}). This research has made 
use of NASA's Astrophysics Data System.

\appendix 
\section{Notes on individual RR Lyrae stars}\label{individualnotes}

{\it V52)} This variable shows up, both in CMDs and in PLs
in all the literature focussed on the RRLs of $\omega$ Cen,
as an overluminous source. As pointed out by \citet{navarrete15} 
this is due to an unresolved blend with a close companion at 
0.5 arcseconds, according to \hst data \citep{anderson10}. 
We have resolved the blend because, both in the 
$K_s$,$J-K_s$ CMD and in the $JHK_s$ PLs, it is well within the distribution
of the RRLs of $\omega$ Cen. This result was achieved by using 
only LCO15 data, since they were collected with very good seeing
(see Table~1).\\ 

{\it V68)} \citet{nemec1994} suggested that this variable is a 
candidate  Anomalous Cepheid (AC). Recently, \citet{navarrete17} claimed 
that it is not possible to discriminate whether V68 is an RRc 
or an AC, because it follows both the $K_s$-band PL relations
for RRc and ACs. They adopted the $K_s$-band PL 
relation for Fundamental (FU) ACs derived by \citet{ripepi2014}
on the basis of ACs in the LMC. However, their relation stops 
at $\sim$0.6 days, because FU ACs in the MCs  
\citep[OGLE,][]{soszynski2015a} have only periods longer than 0.6 days. 
Moreover, there is no theoretical evidence \citep{fiorentino06} 
to support the existence of FU ACs with periods shorter than 0.6 days.
Therefore, the PL that should be adopted is that for 
the FO pulsators \citep{ripepi2014}. At the period of
V68, assuming a distance of $\mu$=13.69 mag (our paper) and 
taking account of the reddening, a FO AC should have a 
$K_s$-band magnitude $\sim$12.46 mag. This is
$\sim$0.45 mag brighter than our value for V68 
($K_s$=12.942$\pm$0.005). This is more than 4$\sigma$ 
away from the PL($K_s$) relation of FO ACs 
\citep[$\sigma_{PLK,AC}$=0.10 mag,][]{ripepi2014}.
Furthermore, no FU AC is known to be this faint, 
also in the optical ($M_V \sim$0.21 mag).
This evidence indicates that V68 is either a background FO 
AC---but this is in contrast with membership probabilities (100\%) derived 
by \citet{vanleeuwen2000} and \citet{bellini2009} 
on the basis of proper motions---or, more probably, 
that it is a member RRc. If we assume that V68 is an RRc, its period 
(0.53462524 days) would be among the largest ever found for an 
RRc in the Halo \citep[GCVs,][]{samus2017} and larger than any of 
the $\sim$11,000 RRc in the Bulge \citep[OGLE,][]{soszynski14} and of the
$\sim$10,000 in the MCs \citep[OGLE,][]{soszynski2016}.\\

{\it V84)} \citet{nemec1994} suggested that this star is a 
candidate AC. Recently, \citet{navarrete17} claimed, 
on the basis of the $K_s$-band PL relations of FU ACs \citep{ripepi2014}, 
that it is not possible to discriminate between a 
foreground RRab and a member AC. For the same reasons already
explained in the note of V68, one should use the relation for FO
pulsators. In case V84 was a cluster-member FO AC, its $K_s$-band 
magnitude should be $\sim$12.32 mag. This is
$\sim$0.45 mag brighter than our estimate (12.781$\pm$0.009 mag).
We rule out the possibility that V84 is a member FO AC. Either
it is a background FO AC or it is a foreground RRab. These 
hypotheses also agree with \citet{vanleeuwen2000}, who  
provides a membership probability of
0\% for V84, on the basis of proper motions. Finally, let us mention that we cannot 
exclude that V84 might be affected by an unresolved blend and new spectroscopic 
measurements of the radial velocity would be highly desirable to assess the 
membership of this variable.
Interestingly enough, the referee drawn our attention on the similarity of 
the light curve of V84 with the variable V70 in M3 
\citep[see Fig.~1 in][]{jurcsik2015}. V70 in M3 is an 
evolved, overluminous RRc variable 
with a period of 0.486 days, that is 0.030 days 
longer than the shortest-period RRab in the cluster.\\

{\it  V159)} The ASAS-SN data are affected by blending, therefore we do not provide
an optical mean magnitude, nor an amplitude in Table~\ref{tab:rrl_asassn}.
The comparison with the literature \citep[][<$m_{Ph}$>=14.68 mag and A{$Ph$}=0.57 
mag, from photographic plates]{vangent1948} gives both a brighter mean magnitude 
(12.395$\pm$0.010 mag) and a smaller amplitude A$V$=0.072$\pm$0.010 mag.\\

{\it V171)} From ASAS-SN optical data, we have derived 
the period of this variable (P=0.52099438 days) 
for the first time. The period and the sawtooth shape of the light curve
tell us that V171 is indeed an RRab star.
We have retrieved 2MASS and VHS data and found that V171
obeys to the PL relation of the RRLs in $\omega$ Cen; we therefore 
suggest that it is a cluster member.\\

{\it V178)} The ASAS-SN $V$-band 
time series, shows no sign of variability. 
This star should be excluded definitively from the list
of candidate RRLs in $\omega$ Cen.\\ 

{\it V179)} From ASAS-SN optical data, we have derived for the first time the 
period of this variable (0.50357939 days).
However, its light curve is that of an
eclipsing binary, with deep minima ($\sim$0.4-0.5 mag) 
in the $V$ band.\\ 

Note that finding charts of V171, V178 and V179 are 
available \citep{wilkens1965}.
Since their coordinates are more than 50 years old, 
we have visually inspected the images from the DSS and the 
finding charts to avoid wrong matches in the modern
ASAS-SN, 2MASS and VHS catalogs.\\

{\it V182)} This star is marked as a ``RR0?'' with a period of 0.5454 days
and coordinates RA=13:32:13.42 and DEC=-47:06:18.6 in the 
\citet{Clement01} catalog. Regrettably, no finding chart is
available and the ASAS-SN light curve of the 
star located at these coordinates shows a constant behavior.
However, a star of similar magnitude is located at RA=13:32:18.71 and 
DEC=-47:06:13.5 (coordinates from the 2MASS Point Source Catalog) and
its light curve is sawtooth-shaped with a period of 0.54539506 days. 
From its period, we conclude that the 
star that we have found is V182 itself, but its
coordinates were wrong. We confirm that V182 is an
RRab star, but it is not a cluster member, because its 
mean $J$, $H$ and $K_s$ mean magnitudes place it $\sim$0.4 mag 
below the PL relation of the RRL stars of $\omega$ Cen.\\

{\it NV411)} This star was missed in our analysis of 
optical data \citep{braga16} due probably to a mismatch
between our point source optical catalog and the catalog of
RRLs of $\omega$ Cen. The successful match between 
the catalog of RRLs and our point source NIR catalog
allowed us to check again the optical data. Finally, we
have retrieved both optical and NIR time series for this star.
The $UBVRI$ mean magnitudes and amplitudes of NV411 are listed 
in Table~\ref{tab:rrl_asassn}.\\

\bibliographystyle{aa}
\bibliography{ms}

\clearpage
\tablenum{1}
\begin{deluxetable*}{lllll rrrr c}
\tablewidth{0pt}
\tabletypesize{\scriptsize}
\tablecaption{Log of the observations of $\omega$ Cen in NIR bands.}
\tablehead{Run & ID & Dates & Telescope & Camera & $J$ & $H$ & $K_s$& other & multiplex \\}
\startdata
     1 & giuseppe   &   2000 Jan 13--2004 Jun 04 & ESO NTT 3.6m  & SOFI     & 109 & \ldots & 130   & \ldots & \ldots\\
     2 & milena     &   2004 Jun 03              & ESO NTT 3.6m  & SOFI     &  15 & \ldots &\ldots & \ldots & \ldots\\
     3 & anna       &   2005 Apr 02              & ESO NTT 3.6m  & SOFI     &  12 &  12    &  36   & \ldots & \ldots\\
     4 & calamid    &   2007 Apr 03--Jun 03      & ESO VLT 8.0m  & MAD      &   6 & \ldots &   9   &  55    & \ldots\\
     5 & blanco     &   2010 May 24--25          & CTIO 4.0m     & Newfirm  &  25 & \ldots &  52   & \ldots & 4     \\
     6 & lco1306    &   2013 Jun 25--29          & Magellan 6.5m & FourStar & 900 & 900    & 900   & \ldots & 4     \\
     7 & lco1501    &   2015 Jan 26--27          & Magellan 6.5m & FourStar & 315 & 315    & 365   & \ldots & 4     \\
\enddata
\tablecomments{\\                                                                    
1 ESO program IDs 64.N-0038(A), 66.D-0557(A), 68.D-0545(A), 073.D-0313(A); observer(s) unknown   \\
2 ESO program ID 073.D-0313(A); observer unknown                                                 \\
3 ESO program ID 59.A-9004(D); observer unknown                                                  \\
4 ESO program ID ``ID96406''; observer unknown; ``other'' = Brackett $\gamma$                \\
5 Proposal ID ``noao''; observers Allen, DePropris                                               \\
6 Observer A. Monson                                                                             \\
7 Observer N. Morrel}                                         
\label{tbl:omegalog_nir}
\end{deluxetable*}
%
 \tablenum{2}
 \setlength{\tabcolsep}{0.05cm}
\begin{deluxetable*}{l cc  l  ccc  ccc  ccc}
 \tablewidth{0pt}
 \tabletypesize{\scriptsize}
 \tablecaption{NIR---$JHK_s$---mean magnitudes and amplitudes for $\omega$ Cen RRLs.}
 \tablehead{  ID & Period (new) & Period (old)\tablenotemark{a} & Variable  &
 fit$J$\tablenotemark{b}& 
 $J$& 
 $AJ$ & 
 fit$H$\tablenotemark{b}& 
 $H$& 
 $AH$ & 
 fit$K_s$\tablenotemark{b}& 
 $K_s$& 
 $AK_s$ \\
  & days & days &type & & mag & mag  & & mag  & mag & & mag & mag }
 \startdata                   
    V3                     & 0.84126158                  &    \ldots                     &  RRab                      &  P  & 13.182$\pm$0.007 &  0.353$\pm$0.029 &   P  & 12.895$\pm$0.007 &  0.300$\pm$0.029 &   S  & 12.864$\pm$0.009 &  0.285$\pm$0.027    \\
    V4                     & 0.62731846                  &    \ldots                     &  RRab                      &  S  & 13.378$\pm$0.008 &  0.494$\pm$0.054 &   S  & 13.106$\pm$0.008 &  0.333$\pm$0.031 &   S  & 13.077$\pm$0.009 &  0.316$\pm$0.020    \\
    V5\tablenotemark{Bl}   & 0.51528002                  &    \ldots                     &  RRab                      &  S  & 13.668$\pm$0.012 &  0.441$\pm$0.028 &   S  & 13.488$\pm$0.019 &  0.354$\pm$0.043 &   S  & 13.409$\pm$0.012 &  0.334$\pm$0.016    \\
    V7                     & 0.71303420                  &    \ldots                     &  RRab                      &  S  & 13.289$\pm$0.005 &  0.428$\pm$0.041 &   P  & 13.018$\pm$0.006 &  0.305$\pm$0.025 &   S  & 12.976$\pm$0.006 &  0.312$\pm$0.028    \\
    V8                     & 0.52132593                  &    \ldots                     &  RRab                      &  S  & 13.577$\pm$0.004 &  0.449$\pm$0.056 &   S  & 13.341$\pm$0.009 &  0.344$\pm$0.018 &   S  & 13.305$\pm$0.005 &  0.295$\pm$0.017    \\
    V9\tablenotemark{Bl}   & 0.52335114\tablenotemark{c}& 0.52346446                     &  RRab                      &  S  & 13.641$\pm$0.010 &  0.320$\pm$0.050 &   P  & 13.430$\pm$0.005 &  0.175$\pm$0.033 &   S  & 13.378$\pm$0.009 &  0.225$\pm$0.025    \\
    V10                    & 0.37475609\tablenotemark{d}& 0.37488161                     &  RRc                       &  P  & 13.551$\pm$0.006 &  0.153$\pm$0.013 &   S  & 13.316$\pm$0.017 &     \ldots       &   S  & 13.313$\pm$0.009 &  0.080$\pm$0.010    \\
\enddata
\end{deluxetable*}

 \tablenum{3}
 \begin{deluxetable*}{l l cc}
 \tablewidth{0pt}
 \tabletypesize{\scriptsize}
 \tablecaption{New optical mean magnitudes and amplitudes of RRL candidates in $\omega$ Cen.}
 \tablehead{ID  & band & mean & Amplitude \\
  & & mag & mag }
 \startdata
  V80\tablenotemark{a}   & $V$  &  14.399$\pm$0.050   & 0.322$\pm$0.050  \\
  V151\tablenotemark{a}  & $V$  &  14.422$\pm$0.050   & 0.236$\pm$0.050  \\
  V159\tablenotemark{b}  & $V$  &   \multicolumn{2}{c}{blended}          \\
  V171\tablenotemark{a}  & $V$  &  14.417$\pm$0.040   & 0.798$\pm$0.043  \\
  V173\tablenotemark{a}  & $V$  &  14.556$\pm$0.050   & 0.455$\pm$0.051  \\
  V175\tablenotemark{a}  & $V$  &  13.254$\pm$0.020   & 0.438$\pm$0.022  \\
  V177\tablenotemark{a}  & $V$  &  14.571$\pm$0.050   & 0.429$\pm$0.061  \\
  V178\tablenotemark{a}  & $V$  &  13.510$\pm$0.024   &        \ldots    \\
\enddata
 \tablecomments{All mean magnitudes and amplitudes were derived from spline fits.}
 \tablenotetext{a}{~From ASAS-SN optical data.}
 \tablenotetext{b}{~The ASAS-SN optical light curve is heavily affected by blending, see
 notes in the individual variables appendix.}
 \tablenotetext{c}{~From our own optical data.}
 \label{tab:rrl_asassn}
 \end{deluxetable*}

\tablenum{4}
\begin{deluxetable*}{l c c c c c c}
\tablewidth{0pt}
\tabletypesize{\scriptsize}
\tablecaption{Mean amplitude ratios of RRc, short-period RRab and long-period
RRab stars of $\omega$ Cen RRLs.}
\tablehead{   
              & \multicolumn{2}{c}{RRc} & \multicolumn{2}{c}{RRab (short)} & \multicolumn{2}{c}{RRab (long)}  \\
              &   mean    &    $\sigma$  &   mean    &  $\sigma$ &   mean    &  $\sigma$       }
\startdata                      
  A$J$/A$B$  & 0.316 & 0.062 & 0.350 & 0.063 & 0.394 & 0.097 \\
  A$H$/A$B$  & 0.198 & 0.058 & 0.237 & 0.055 & 0.331 & 0.161 \\
  A$K_s$/A$B$ & 0.172 & 0.051 & 0.233 & 0.034 & 0.323 & 0.126 \\
  A$J$/A$V$  & 0.396 & 0.066 & 0.429 & 0.057 & 0.488 & 0.106 \\
  A$H$/A$V$  & 0.252 & 0.071 & 0.286 & 0.055 & 0.414 & 0.202 \\
  A$K_s$/A$V$ & 0.221 & 0.061 & 0.280 & 0.028 & 0.407 & 0.169 \\
  A$H$/A$J$   & 0.647 & 0.183 & 0.676 & 0.118 & 0.861 & 0.238 \\
  A$K_s$/A$J$ & 0.551 & 0.128 & 0.658 & 0.093 & 0.822 & 0.229
\enddata
\label{tab:amplratio}
\end{deluxetable*}

\tablenum{5}
\begin{deluxetable*}{l r r r r r r r r r}
\tablewidth{0pt}
\tabletypesize{\scriptsize}
\tablecaption{Coefficients of the observed PL relations of RRab, RRc and 
RRab plus fundamentalized RRc stars of $\omega$ Cen of the 
form $X$ = $a$ + $b\log{P}$.}
\tablehead{  & \multicolumn{3}{c}{RRc} & \multicolumn{3}{c}{RRab} & \multicolumn{3}{c}{Global} \\
 & $a$ & $b$ & $\sigma$ & $a$ & $b$ & $\sigma$ & $a$ & $b$ & $\sigma$ \\
 & mag & mag & mag & mag & mag  & mag  & mag & mag & mag }
\startdata                      
$J$  &      12.666 &      --2.105 &  0.036 &      12.963 &      --2.318 &  0.050 &      13.018 &      --1.884 &  0.043 \\
     &  $\pm$0.030 & $\pm$ 0.067 &        & $\pm$0.014 & $\pm$0.073 &        & $\pm$0.010 & $\pm$0.034 & \\
$H$  &      12.314 &      --2.403 &  0.040 &      12.643 &      --2.520 &  0.046 &      12.686 &      --2.224 &  0.043 \\
     & $\pm$0.034 & $\pm$0.075 &        & $\pm$0.014 & $\pm$0.069 &        & $\pm$0.009 & $\pm$0.034 & \\
$K_s$&      12.244 &      --2.531 &  0.042 &      12.591 &      --2.621 &  0.048 &      12.625 &      --2.380 &  0.046 \\
     & $\pm$0.035 & $\pm$0.076 &        & $\pm$0.014 & $\pm$0.070 &        & $\pm$0.010 & $\pm$0.036 & \\
\enddata
\label{tab:pl}
\end{deluxetable*}

\tablenum{6}
\begin{deluxetable*}{l r r r r r r r r r r r r}
\tablewidth{0pt}
\tabletypesize{\scriptsize}
\tablecaption{Coefficients of the observed PLZ relations of RRab, RRc and 
RRab plus fundamentalized RRc stars of $\omega$ Cen of the 
form $X$ = $a$ + $b\log{P}$ + $c$[Fe/H].}
\tablehead{  & \multicolumn{4}{c}{RRc} & \multicolumn{4}{c}{RRab} & \multicolumn{4}{c}{Global} \\
 & $a$ & $b$ & $c$ & $\sigma$ & $a$ & $b$ & $c$ & $\sigma$ & $a$ & $b$ & $c$ & $\sigma$ \\
 & mag & mag & mag & mag & mag  & mag  & mag & mag & mag & mag & mag & mag \\}
\startdata                      
\multicolumn{13}{c}{---[Fe/H] from \citet{rey2000}---} \\
\multicolumn{13}{c}{---Free parameters---} \\
$J$&      12.711 &      --2.112      &  0.026 &  0.034 &      13.093 &      --2.284 &  0.077 &  0.038 &      13.124 &      --1.873 &  0.061 &  0.046 \\
   & $\pm$0.058 & $\pm$0.092 & $\pm$0.018 &       & $\pm$0.035 & $\pm$0.071 & $\pm$0.018 &       & $\pm$0.032 & $\pm$0.052 & $\pm$0.016 & \\
$H$&      12.374 &      --2.394      &  0.034 &  0.036 &      12.780 &      --2.502 &  0.081 &  0.034 &      12.779 &      --2.225 &  0.056 &  0.040 \\
   & $\pm$0.061 & $\pm$0.097 & $\pm$0.019 &       & $\pm$0.030 & $\pm$0.063 & $\pm$0.015 &       & $\pm$0.028 & $\pm$0.045 & $\pm$0.014 & \\
$K_s$&    12.304 &      --2.531      &  0.033 &  0.038 &      12.753 &      --2.601 &  0.096 &  0.038 &      12.725 &      --2.386 &  0.062 &  0.045 \\
   & $\pm$0.065 & $\pm$0.103 & $\pm$0.021 &       & $\pm$0.033 & $\pm$0.069 & $\pm$0.017 &       & $\pm$0.030 & $\pm$0.050 & $\pm$0.015 & \\
\enddata
\label{tab:plz}
\tablenotetext{a}{We have fixed $b$ at the value of the
slope of the predicted PLZ.}
\end{deluxetable*}

\clearpage

\tablenum{7}
\begin{deluxetable*}{l rrr  rrr  rrr }
\tablewidth{0pt}
\tabletypesize{\scriptsize}
\tablecaption{True distance moduli to $\omega$ Cen estimated using both the theoretical and the 
empirical calibration of NIR PLZ relations for RRLs.}
\tablehead{  &  \multicolumn{3}{c}{RRc} &  \multicolumn{3}{c}{RRab} &  \multicolumn{3}{c}{Global} \\
 & DM & err\tablenotemark{a} & $\sigma$\tablenotemark{b} & DM & err\tablenotemark{a} & $\sigma$\tablenotemark{b} & DM & err\tablenotemark{a} & $\sigma$\tablenotemark{b} \\
 & mag & mag & mag & mag & mag  & mag  & mag & mag & mag }
\startdata
\multicolumn{10}{c}{---[Fe/H] from \citet{sollima06a}---} \\
\multicolumn{10}{c}{---Theoretical calibration---} \\
         $J$   & 13.658 &  0.009 &  0.047 & 13.656 &  0.012 &  0.029 & 13.655 &  0.009 &  0.035 \\
         $H$   & 13.675 &  0.006 &  0.044 & 13.676 &  0.008 &  0.034 & 13.674 &  0.006 &  0.033 \\
         $K_s$ & 13.697 &  0.005 &  0.041 & 13.687 &  0.007 &  0.026 & 13.690 &  0.005 &  0.029 \\                                                              
mean$_{JHK_s}$ & 13.677 &  0.007 &  0.047 & 13.673 &  0.009 &  0.032 & 13.673 &  0.007 &  0.036 \\                                                              
\multicolumn{10}{c}{{\bf Overall mean: 13.674$\pm$0.008$\pm$0.038}} \\
\multicolumn{10}{c}{---Empirical calibration---} \\
\enddata
\tablenotetext{a}{Uncertainty on the DM derived from the propagation of the 
photometric error, uncertainties on the coefficients of the PLZ and uncertainties 
on the iron abundances.}
\tablenotetext{b}{Standard deviation of the DMs of single RRLs}
\label{tab:dmod}
\end{deluxetable*}

\clearpage

\tablenum{8}
\begin{deluxetable*}{l l l c}
\tablewidth{0pt}
\tabletypesize{\scriptsize}
\tablecaption{True distance moduli to $\omega$ Cen: literature and current estimates.}
\tablehead{$\mu$ & E(\bmv) & Ref. & Notes\tablenotemark{a} \\ mag & mag & & }
\startdata
\multicolumn{4}{c}{---Geometrical (13.56$\pm$0.14 mag)---} \\
13.65$\pm$0.12                                                & \ldots        & \citet{thompson2001} & (1)  \\
13.75$\pm$0.04                                & 0.11\tablenotemark{b}         & \citet{kaluzny2002} & (2) \\
13.41$\pm$0.14                                                & \ldots        & \citet{vandeven2006} &   (3)  \\
13.49$\pm$0.14~;~13.51$\pm$0.12                               & 0.131         & \citet{kaluzny07a} & (4) \\
\multicolumn{4}{c}{---Optical (13.66$\pm$0.07 mag)---} \\
13.36$\pm$0.10                                                & 0.11          & \citet{cannon1974}    &  (5) \\
13.53$\pm$0.20                                                & 0.11          & \citet{nemec1994}    &  (6) \\
\enddata
\end{deluxetable*}


\end{document}